\documentclass[%
 reprint,
superscriptaddress,
 amsmath,amssymb,
 aps,
]{revtex4-1}

\usepackage{graphicx}
\usepackage{dcolumn}
\usepackage{bm}
\usepackage{color}

\newcommand{\GaVS}{GaV$_4$S$_8$}
\newcommand{\GeVS}{GeV$_4$S$_8$}
\newcommand{\GaMoS}{GaMo$_4$S$_8$}
\newcommand{\GaNbS}{GaNb$_4$S$_8$}
\newcommand{\AlVS}{AlV$_4$S$_8$}
\newcommand{\AlMoS}{AlMo$_4$S$_8$}
\newcommand{\GaVSe}{GaV$_4$Se$_8$}
\newcommand{\GaMoSe}{GaMo$_4$Se$_8$}
\newcommand{\GaTaSe}{GaTa$_4$Se$_8$}
\newcommand{\GaNbSe}{GaNb$_4$Se$_8$}
\newcommand{\TN}{\ensuremath{T_{\mathrm{N}}}}
\newcommand{\TC}{\ensuremath{T_{\mathrm{C}}}}
\newcommand{\TJT}{\ensuremath{T_{\mathrm{JT}}}}

\begin{document}

\title{Lattice dynamics and electronic excitations in a large family of lacunar spinels with a breathing pyrochlore lattice structure}
\date{\today}

\author{S.~Reschke}
\affiliation{Experimentalphysik V, Center for Electronic
Correlations and Magnetism, Institute of Physics, Augsburg
University, D-86159 Augsburg, Germany}

\author{F.~Meggle}
\affiliation{Experimentalphysik II, Institute of Physics, Augsburg
University, D-86159 Augsburg, Germany}

\author{F.~Mayr}
\affiliation{Experimentalphysik V, Center for Electronic
Correlations and Magnetism, Institute of Physics, Augsburg
University, D-86159 Augsburg, Germany}

\author{V.~Tsurkan}
\author{L.~Prodan}
\affiliation{Experimentalphysik V, Center for Electronic
Correlations and Magnetism, Institute of Physics, Augsburg
University, D-86159 Augsburg, Germany} \affiliation{Institute of
Applied Physics, MD-2028~Chi\c{s}in\u{a}u, Republic of Moldova}

\author{H.~Nakamura}
\affiliation{Department of Materials Science and Engineering, Kyoto University, Kyoto 606-8501, Japan}

\author{J.~Deisenhofer}
\affiliation{Experimentalphysik V, Center for Electronic
Correlations and Magnetism, Institute of Physics, Augsburg
University, D-86159 Augsburg, Germany}

\author{C.~A.~Kuntscher}
\affiliation{Experimentalphysik II, Institute of Physics, Augsburg
University, D-86159 Augsburg, Germany}

\author{I.~K\'{e}zsm\'{a}rki}
\affiliation{Experimentalphysik V, Center for Electronic
Correlations and Magnetism, Institute of Physics, Augsburg
University, D-86159 Augsburg, Germany}

\date{February 3, 2020}

\begin{abstract}
Reproducing the electronic structure of AM$_4$X$_8$ lacunar spinels with a breathing pyrochlore lattice is a great theoretical challenge due to the interplay of various factors. The character of the M$_4$X$_4$ cluster orbitals is critically influenced by the Jahn-Teller instability, the spin-orbit interaction, and also by the magnetic state of the clusters. Consequently, to reproduce the narrow-gap semiconducting nature of these moderately correlated materials requires advanced approaches, since the strength of the inter-cluster hopping is strongly affected by the character of the cluster orbitals. In order to provide a solid experimental basis for theoretical studies, we performed broadband optical spectroscopy on a large set of lacunar spinels, with systematically changing ions at the A and M sites as well as the ligand (A=Ga, Ge, Al; M=V, Mo, Nb, Ta; X=S, Se). Our study covers the range of phonon excitations and also electronic transitions near the gap edge. In the phonon excitation spectrum a limited subset of the symmetry allowed modes is observed in the cubic state, with a few additional modes emerging upon the symmetry-lowering structural transition. All the infrared active modes are assigned to vibrations of the ligands and ions at the A sites, with no obvious contribution from the M-site ions. Concerning the electronic states, we found that all compounds are narrow-gap semiconductors ($E_\mathrm{g} = 130 - 350$\,meV) already in their room-temperature cubic state and their structural transitions induce weak, if any, changes in the band gap. The gap value is decreased when substituting S with Se and also when replacing $3d$ ions by $4d$ or $5d$ ions at the M sites.
\end{abstract}

\maketitle

\section{Introduction}

\begin{figure}[tb]
\includegraphics[width = 1\columnwidth]{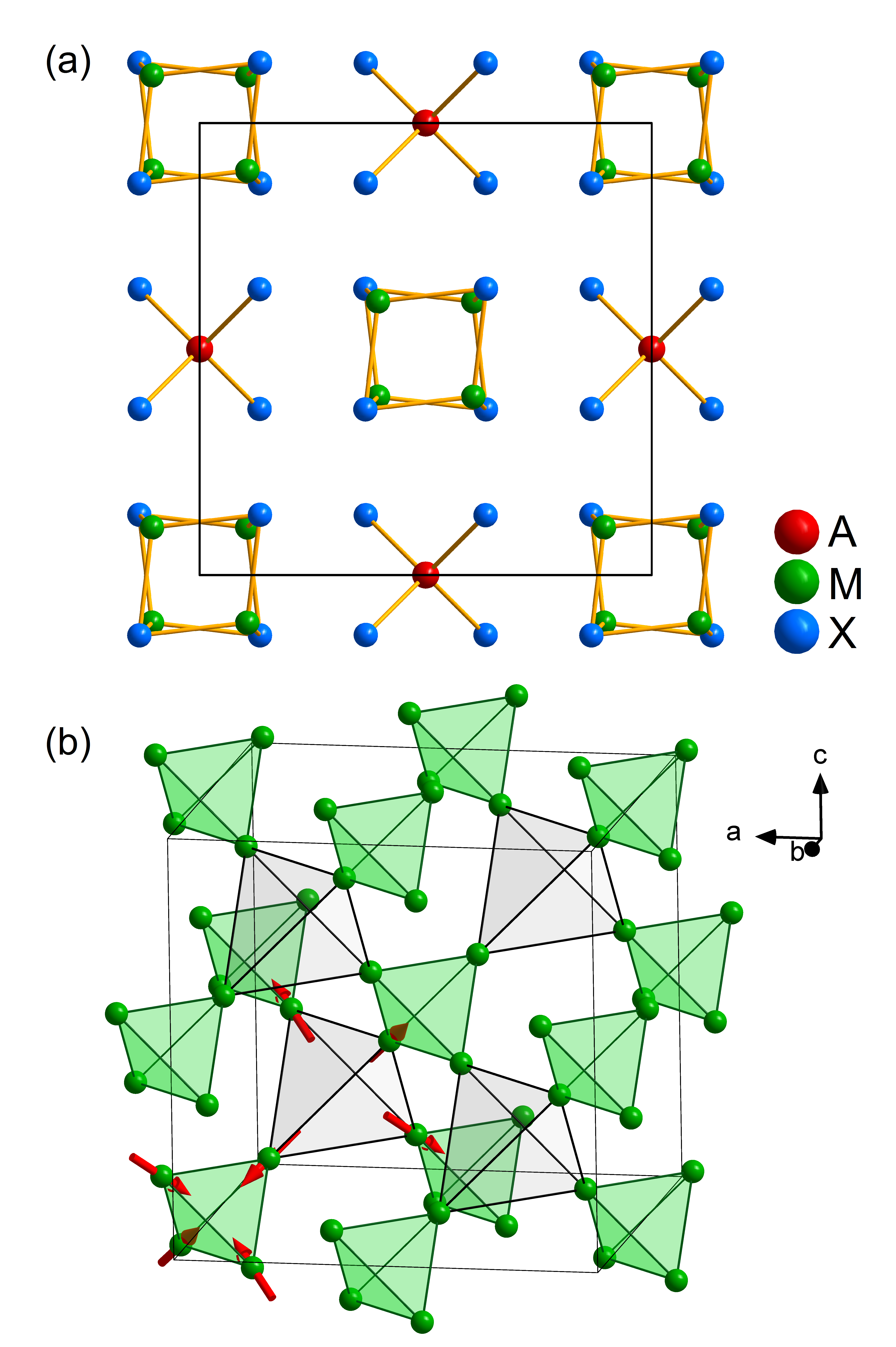}
\caption{\label{fig:Crystal_structure} (a) Crystallographic structure of the cubic high-temperature phase of the lacunar spinels AM$_4$X$_8$, as viewed along the [100] direction, consisting of M$_4$X$_4$ cubane clusters and AX$_4$ tetrahedra. The two types of polyhedra form two fcc lattices shifted relative to each other. (b) M sublattice forming a breathing pyrochlore lattice. The arrows indicate the shift of the M sites with respect to an ideal pyrochlore lattice.}
\end{figure}

The investigated compounds belong to the family of the lacunar spinels AM$_4$X$_8$ (A=Ga, Ge, Al; M=V, Mo, Nb, Ta; X=S, Se) \cite{Barz:1973,BenYaich:1984,Johrendt:1998,Pocha:2000}. The crystal structure of the lacunar spinels is derived from a normal spinel by removing every second atom, in a regular fashion, from the diamond lattice formed by the A sites. Owing to the ordered voids, the pyrochlore sublattice of M sites develops a breathing, i.e., it becomes a network of larger and smaller M$_4$ tetrahedra alternating in a regular fashion \cite{Kezsmarki:2015,Butykai:2017} as illustrated by Fig. \ref{fig:Crystal_structure}(b). This results in a non-centrosymmetric cubic crystal structure with space group $F\overline{4}3m$ at room temperature, consisting of two fcc lattices, one with weakly linked M$_4$X$_4$ molecular clusters and the other with AX$_4$ clusters, \cite{Barz:1973,Johrendt:1998,Vandenberg:1975} as shown in Fig. \ref{fig:Crystal_structure}(a).

The M$_4$ tetrahedral clusters are the magnetic building blocks of the lacunar spinels and can be described by a molecular orbital scheme \cite{Harris:1989}. In this picture, the highest occupied cluster orbital is triply degenerate. Due to its partial occupation, the M$_4$X$_4$  clusters in the investigated lacunar spinels are Jahn-Teller active and, thus, subject to structural distortion. Consequently, the orbital degeneracy is lifted by a symmetry lowering cooperative Jahn-Teller distortion in the temperature range between 30 and 50\,K, depending on the compounds. This results in rhombohedral $R3m$ (\GaVS{}, \GaMoS{}, \GaVSe{}, \GaMoSe{}, \AlMoS{}) \cite{Pocha:2000,Mueller:2006,Francois:1991,Fujima:2017,Francois:1992}, tetragonal $P\overline{4}2_1m$ (\GaNbS{}, \GaTaSe{}) \cite{Jakob:2007,Jakob:2007a}, or orthorhombic $Imm2$ (\GeVS{}) \cite{Mueller:2006,Bichler:2008} low temperature symmetries.

Several of the lacunar spinels also undergo magnetic ordering at lower temperatures, hosting various magnetic states such as the cycloidal state and the N\'{e}el-type skyrmion lattices in \GaVS{}, \GaVSe{} and \GaMoS{} \cite{Kezsmarki:2015,Bordacs:2017,Butykai:2019}. Dielectric spectroscopy data imply that the cubic to rhombohedral order in \GaVS{}, \GaVSe{} and \GaMoS{} is a disorder to order type, implying that the dynamic Jahn-Teller distortion is present already at temperatures well above the cooperative long-range order \cite{Wang:2015,Ruff:2017,Geirhos:2018}. Table \ref{tab:lacSpinels} summarizes the structural and magnetic phase transition temperatures, low-temperature structures and type of magnetic ordering of the lacunar spinels investigated in this work.

A plethora of physical phenomena emerging from correlations and spin-orbit interaction in lacunar spinels have attracted high interest from both the experimental and theoretical sides: Pressure-induced superconductivity \cite{Abd:2004,Pocha:2005}, bandwidth-controlled metal-to-insulator transitions \cite{Phuoc:2013,Camjayi:2014}, resistive switching \cite{Cario:2010,Dubost:2013,Stoliar:2013}, large negative magnetoresistance \cite{Dorolti:2010}, a two-dimensional topological insulating state \cite{Kim:2014} and orbital order induced ferroelectricity \cite{Singh:2014}. However, to reproduce the electronic structure of the lacunar spinels, especially in the cubic state, is still a great challenge for theory. Due to fractional occupation of the M$_4$X$_4$ cluster orbitals one would naively expect a metallic behavior. In contrast, most of the lacunar spinels are semiconducting, pointing toward the relevance of electron-electron correlations \cite{Shanthi:1999}.

According to theoretical studies, standard density functional theory (DFT) calculations fail to open the band gap experimentally observed in the cubic phase of lacunar spinels, instead they give rise to a metallic state \cite{Johrendt:1998,Shanthi:1999,Pocha:2005,Camjayi:2012,Cannuccia:2017,Kim:2018,Wang:2019}. The common aspect of all studies performed within the DFT+$U$ scheme is that besides finite Coulomb interactions ($U$) they require a lattice symmetry lower than cubic in order to open a gap \cite{Pocha:2005,Mueller:2006,Jakob:2007,Cannuccia:2017,Wang:2019}. Indeed, DFT studies of the $R3m$ phase emphasize the importance of lattice distortions on the electronic structure of the lacunar spinels: Structural distortions alone, i.e., without finite $U$, have been found to be sufficient to open a gap at $E_\mathrm{F}$ in \GaMoS{}, \GaVS{}, \GaMoSe{} and \GaVSe{} \cite{Sieberer:2007,Wang:2019}.

A recent theoretical work on \GaVS{} \cite{Kim:2018} suggests, that molecular orbital cluster dynamical mean field theory (cluster DMFT) is a promising tool for the description of the electronic structure of lacunar spinels and can reproduce the semiconducting nature of the material with realistic values of $U$, if a sufficiently large number of cluster orbitals are included in the calculation. Similar results were reported for \GaTaSe{}, where DMFT with $U$ in the range of 1\,eV could open a gap and also reproduce the temperature-induced metal to insulator transition \cite{Camjayi:2014}. Recent calculations systematically assessing the performance of various exchange-correlation potentials could reproduce a finite gap in the cubic phase in \GaVS{}, \GaMoS{}, \GaNbSe{} and \GaTaSe{} using the PBE functional only including spin orbit coupling in addition to a Hubbard $U$ of 2 to 3\,eV \cite{Wang:2019}. This finding shows that lacunar spinels may be classified as Hund insulators.

The challenge in modeling the electronic structure of lacunar spinels comes from the interplay of the following key factors: 1) the hopping between the M$_4$X$_4$  clusters is comparable, though somewhat weaker than the intra-cluster $U$ due to the strong breathing of the pyrochlore lattice, 2) the hopping between the clusters critically depends on the form of the cluster orbitals, 3) the character of the cluster orbitals is strongly affected by the strength of the spin-orbit coupling as well as by the structural deformation of the cluster due to the Jahn-Teller activity and 4) the magnetic ground state has also a strong impact on the electronic states. The last point is best exemplified in, e.g., in \GaNbS{}, where the cubic-to-tetragonal structural transition is accompanied with the formation of a spin singlet state \cite{Jakob:2007,Waki:2010}. In this case it is not even clear if the primary drive of the magneto-structural transition is the spin singlet formation or the orbital order.

\begin{table*}[tb]
\caption{\label{tab:lacSpinels}
Structural transition temperature \TJT{}, low-temperature structure, magnetic phase transition temperature \TC{} and type of magnetic order in lacunar spinel compounds. (FM = ferromagnetic, AFM = antiferromagnetic, cyc = cycloidal, SkL = skyrmion lattice).}
\begin{ruledtabular}
\begin{tabular}{lccccc}
 & \TJT{} (K) &  low $T$ structure & \TC{} (K)  & magnetic order &  Literature \\
\hline
\GaVS{}     & 44 &  $R3m$            & 13  &  cyc, SkL, FM  &\cite{Kezsmarki:2015,Widmann:2016a}  \\
\GeVS{}     & 30 &  $Imm2$    &15          & AFM   & \cite{Johrendt:1998,Mueller:2006,Bichler:2008,Widmann:2016}\\
\GaMoS{}    & 45              &  $R3m$               & 20  &  cyc, SkL, FM& \cite{Rastogi:1983,BenYaich:1984,Francois:1991,Butykai:2019} \\
\GaNbS{}    & 32 & $P\overline{4}2_1m$ & 32 &  spin singlet & \cite{Jakob:2007,Tabata:2010,Waki:2010}  \\
\AlVS{}     &   10    &  $R3m$               & 40 & FM & \cite{Bichler:2010}  \\
\AlMoS{}    &  45     &  $R3m$               & 12 & FM &   \cite{Francois:1992,Ikeno:2007} \\
\GaVSe{}    &  41     &  $R3m$               &18& cyc, SkL, FM & \cite{Bordacs:2017,Fujima:2017}   \\
\GaMoSe{}   &  54     &  $R3m$               & 25 & FM &  \cite{Rastogi:1983,Rastogi:1987,BenYaich:1984,Francois:1992} \\
\GaTaSe{}   & 53    &  $P\overline{4}2_1m$ & 53   &  spin singlet  & \cite{Jakob:2007a,Kawamoto:2016}\\
\end{tabular}
\end{ruledtabular}
\end{table*}

The fact that the overall exchange coupling, as measured by the Curie-Weiss constant, changes sign upon the cubic-to-rhombohedral distortion, e.g., in \GaVS{} \cite{Nakamura:2005,Widmann:2016a}, also speaks for the sensitivity of hopping and magnetic exchange paths on the precise form of the cluster orbitals. The possibility to turn the Mott insulator state to a spin-orbit (or Hund) insulator was also discussed, when replacing $3d$ by $4d$ and $5d$ ions at the M sites \cite{Kim:2014}. The large variety of magnetic states reported in lacunar spinels, including ferromagnetic, antiferromagnetic, cycloidal, N\'{e}el-type skyrmion lattice and singlet states \cite{Kezsmarki:2015,Bordacs:2017,Butykai:2019,Leonov:2017,Leonov:2017a,Waki:2010,Tabata:2010,Kawamoto:2016} also points toward the delicate balance of magnetic interactions and their strong dependence on the electronic structure. In this respect, band structure calculations are also facing to a great challenge. The modeling of these magnetic phases requires information about the spin distribution over the M$_4$ clusters and about all types of magnetic interactions in both the cubic and the lower-symmetry states: The exchange couplings and their anisotropy, the Dzyaloshinskii-Moriya interaction, the local anisotropy arising on the M$_4$ clusters, as well as the g-factor and its anisotropy. Moreover, the observation of a strong magnetoelectric effect in these compounds \cite{Ruff:2015,Ruff:2017,Singh:2014} also motivated advanced ab initio calculations of the magnetoelectric properties and multi-scale modeling of magnetic properties \cite{Xu:2015,Zhang:2017,Nikolaev:2019,Zhang:2019,Kitchaev:2019}.

Despite the success of these works of even being able to reproduce the N\'{e}el-type skyrmion lattice state, emerging in some of these compounds, based on magnetic interaction parameters obtained via DFT+$U$ schemes, the question is whether these magnetic parameters can be taken seriously on the quantitative level, given that this approach fails in opening a gap in the cubic state. A certainly attractive scheme to address the proper form of cluster orbitals and magnetic interactions would be the quantum chemistry approach successfully applied for materials such as possible Kitaev-physics hosts \cite{Katukuri:2014,Yadav:2016} and layered copper oxides \cite{Hozoi:2011}.

In order to provide a solid experimental basis, supporting theoretical studies on this compound family, we performed a systematic study of the lattice dynamics as well as the electronic excitations by broadband optical spectroscopy on a wide range of compounds, namely \GaVS{}, \GeVS{}, \GaMoS{}, \GaNbS{}, \AlVS{}, \AlMoS{}, \GaVSe{}, \GaMoSe{} and \GaTaSe{}.  By the sequential replacements of ions on the A and M sublattices as well as the ligand we could assign the infrared-active phonon modes to the different structural units and also trace substitution effects on the gap size.

\section{Experimental Details}

Polycrystalline samples of AM$_4$X$_8$ spinels were prepared by solid state reactions using high-purity elements (Ga 99.9999\,\%, Ge 99.99\,\%, Al 99.95\,\% , V 99.5\,\%, Mo 99.95\,\%, Nb 99.8\,\%, Ta 99.98\,\%, S 99.999\,\%, Se 99.999\,\%) by three repeated synthesis at 800\,$^\circ$C - 1050\,$^\circ$C. After each step of synthesis the phase content was checked by x-ray powder diffraction. The single crystals were grown by the chemical transport method using iodine as transport agent.

Optical reflectivity measurements on \GaVS{}, \GeVS{}, \GaMoS{}, \GaNbS{}, \AlVS{} and \GaVSe{} were performed on as-grown (111) surfaces of single crystals. For single crystals of limited size, the reflectivity spectra were measured using micro-spectroscopy. In the case of \GaNbS{}, where the absolute value of the reflectivity determined by micro-spectroscopy was not fully reliable, we also measured the reflectivity spectrum in the far infrared (IR) range on a large-size high-density polycrystalline pellet, in order to determine the precise absolute value of the reflectivity. For those compounds, where single crystals of sufficient size or high-density pellets were not available (\AlMoS{}, \GaMoSe{}, \GaTaSe{}), transmission was measured on polycrystalline powder samples embedded in polyethylene (PE). 

For optical reflectivity and transmission measurements, the Bruker Fourier-transform IR-spectrometers IFS 113v and IFS 66v/S were used. In the case of crystal sizes below 1\,mm, the measurements were carried out on a Bruker Vertex 80v spectrometer coupled to a Hyperion IR microscope. In these reflectivity studies, the frequency range from 100 to 23000\,cm$^{-1}$ was covered. All spectrometers were equipped with He-flow cryostats (CryoVac), which allowed for measurements at temperatures between 10 and 300\,K. 

The broadband optical conductivity spectra have been calculated from the reflectivity spectra via Kramers-Kronig transformation with a $\omega^{-1.5}$ high-frequency extrapolation, followed by a $\omega^{-4}$ extrapolation above 800000\,cm$^{-1}$. The complex dielectric function $\epsilon(\omega)=\epsilon_1(\omega)-i\epsilon_2(\omega)$ in the far-infrared (FIR) range was obtained from the reflectivity spectra by the Kramers-Kronig constrained variational dielectric function (VDF) method developed by Kuzmenko \cite{Kuzmenko:2005}, implemented in the RefFIT program \cite{Kuzmenko:2018}.

\section{Experimental Results and Discussion}

\begin{figure}[tb]
\includegraphics[width = 0.85\columnwidth]{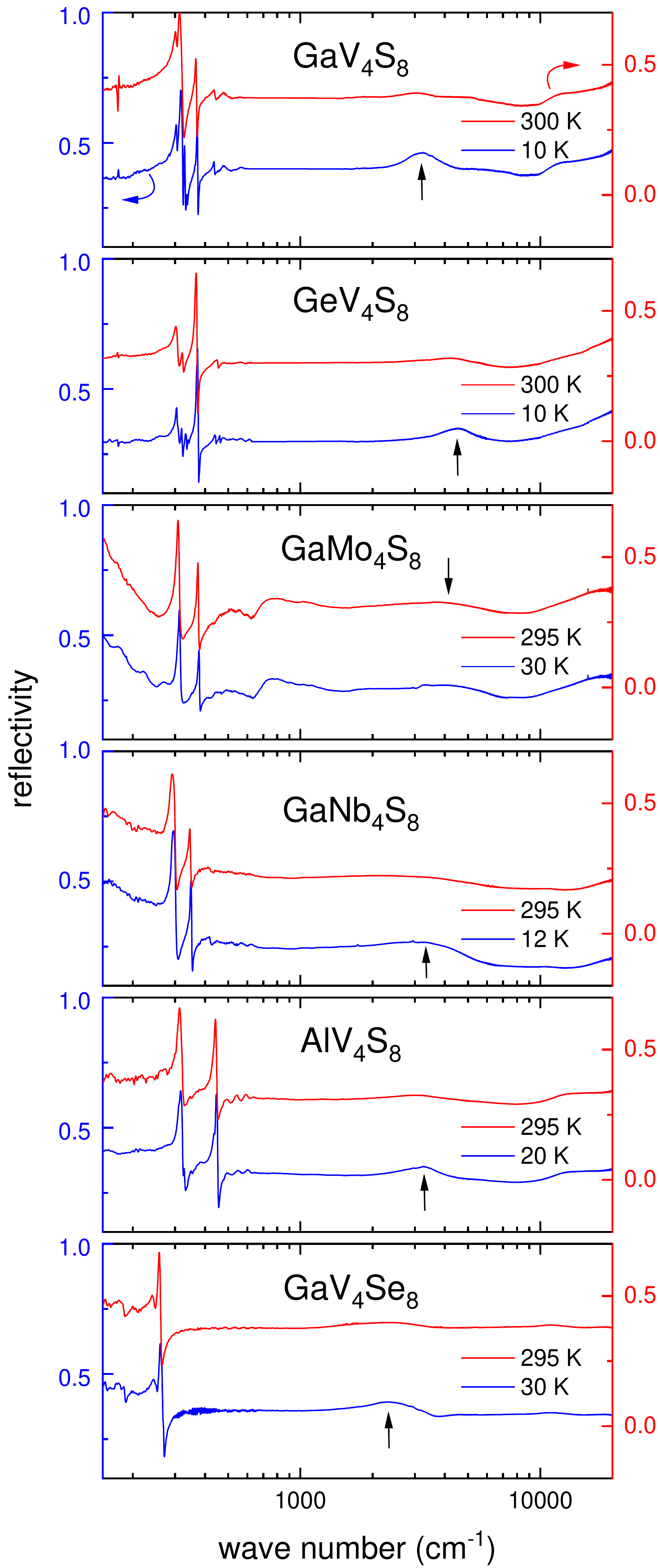}
\caption{\label{fig:broadband_R} Semilogarithmic plot of the reflectivity spectra of \GaVS{}, \GeVS{}, \GaMoS{}, \GaNbS{}, \AlVS{} and \GaVSe{} between 150 and 20000\,cm$^{-1}$, for the low-temperature Jahn-Teller distorted phase (blue curves) and the cubic high-temperature phase (red curves). The manifestation of the band gap as a broad hump in the reflectivity spectra is indicated by arrows.}
\end{figure}

Figure \ref{fig:broadband_R} shows a comparative plot of the frequency-dependent broadband reflectivity spectra of \GaVS{}, \GeVS{}, \GaMoS{}, \GaNbS{}, \AlVS{} and \GaVSe{} over the range of 150 - 20000\,cm$^{-1}$. Data for \GaVS{} and \GeVS{} are reproduced from Ref. \cite{Reschke:2017}. For each compound one spectrum of the cubic room-temperature phase ($F\overline{4}3m$) and one spectrum of the Jahn-Teller distorted low-temperature phase ($R3m$, $Imm2$, or $P\overline{4}2_1m$) is displayed. For all lacunar spinel compounds under investigation, the broadband reflectivity spectra are qualitatively similar. In the frequency range up to 600\,cm$^{-1}$ the phonon resonances can clearly be observed, though the number of the observed modes is less than expected from the analysis of the eigenmodes, as discussed later. For frequencies between 2000 and 6000\,cm$^{-1}$ the reflectivity of each compound exhibits a rather broad excitation feature (arrows in Fig. \ref{fig:broadband_R}), which we interpret here for simplicity as an electronic band gap irrespective of the details of the electronic band structure. This feature becomes more pronounced when lowering the temperature. The peculiar aspect of \GaMoS{} and \GaNbS{}, distinguishing them from the other compounds, is the weak low-frequency upturn of the reflectivity, taking place below  300 - 400\,cm$^{-1}$. The gradual increase of the reflectivity in this low-frequency range is found to be reproducible, but its origin is not clear at present. The origin of a dip observed in the reflectivity of \GaMoS{} at around 700\,cm$^{-1}$ remains also unclear so far.

\subsection{\label{sec:phonons}Phonons}

\begin{figure}[h!tb]
\includegraphics[width = 0.77\columnwidth]{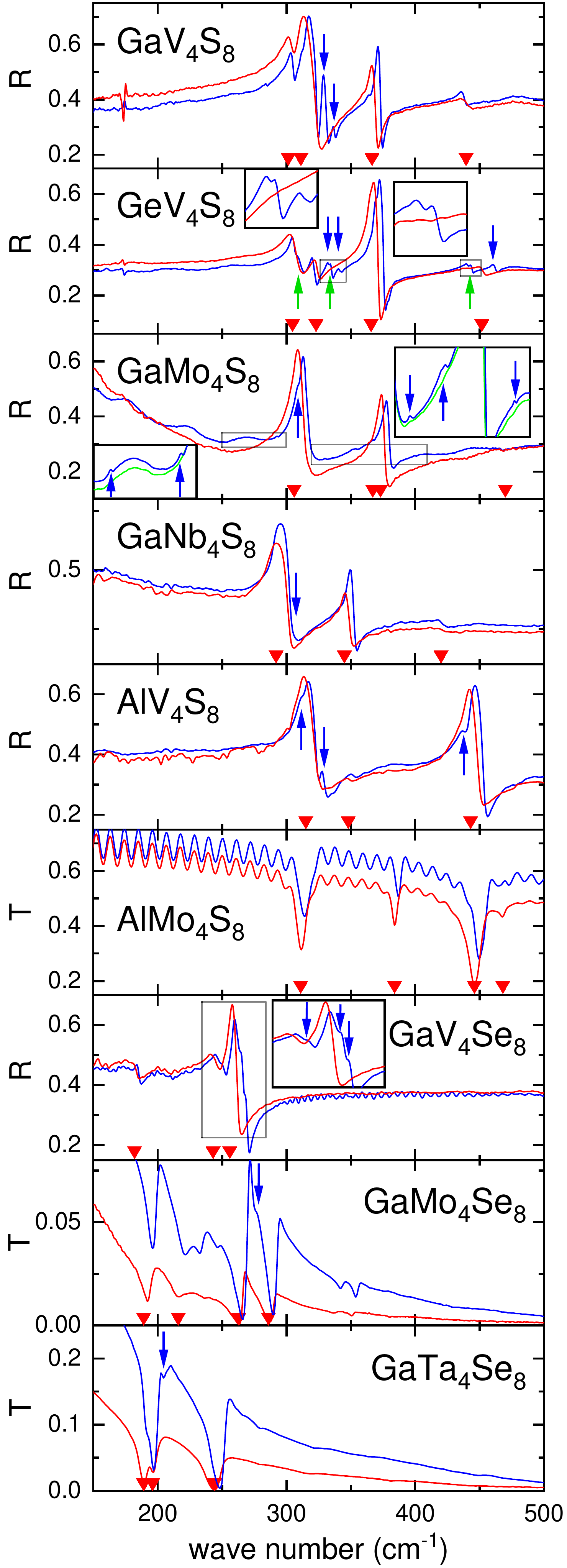}
\caption{\label{fig:FIR_R} Frequency-dependent FIR phonon reflectivity or transmission spectra of the lacunar spinels at room temperature (red curves) and 10\,K (blue curves). The green curve in \GaMoS{} was measured at 50\,K. Reflectivity was measured on single-crystalline samples, transmission data were obtained on polycrystalline samples. Red triangles indicate the experimental phonon eigenfrequencies at room temperature. Blue arrows indicate additional modes appearing at the structural phase transition, green arrows indicate modes appearing at the magnetic phase transition. In the case of weak modes, insets show a zoom-in.}
\end{figure}

A comparative study of the phonon spectra measured in the far-infrared (FIR) range for \GaVS{}, \GeVS{}, \GaMoS{}, \GaNbS{}, \AlVS{} and \GaVSe{} single crystals is shown in Fig. \ref{fig:FIR_R} for 10\,K (blue spectra) and room temperature (red spectra). In addition to the reflectivity spectra of these compounds, FIR transmission spectra obtained on \AlMoS{}, \GaMoSe{} and \GaTaSe{} polycrystalline samples are also displayed in Fig. \ref{fig:FIR_R}.

Group theory allows us to determine the allowed zone center optical phonon modes directly from crystal symmetry and the occupied Wyckoff positions. The lacunar spinels AM$_4$X$_8$ under consideration have cubic crystal symmetry at room temperature with space group $F\overline{4}3m$. For this high-temperature structure, group theory predicts
$\Gamma_\mathrm{M} = \Gamma_{\mathrm{X}^1} = \Gamma_{\mathrm{X}^2} = A_1 + E + F_1 + 2F_2$
modes for the M site and the two different ligand sites (X$^1$ and X$^2$) and
$\Gamma_{\mathrm{A}} = F_2$
for the A site. Thus, the total number of modes of the lacunar spinels in the high-temperature cubic state is
$\Gamma_{\mathrm{AM}_4\mathrm{X}_8} = 3A_1 + 3E + 3F_1 + 7F_2$.
From these modes only the $F_2$ phonon modes are IR active, whereby one $F_2$ mode corresponds to the three acoustic modes. Therefore, a total number of six zone-center IR-active phonon modes can be expected for the high-temperature cubic phase, each of them being triply degenerate. For temperatures below the structural phase transition, the number of allowed IR active phonon modes increases due to the lowering of the crystal symmetry. Not counting the acoustic modes, a total number of 21 IR-active modes can be expected in the rhombohedral phase $R3m$ (\GaVS{}, \GaMoS{}, \GaVSe{}, \GaMoSe{}, \AlMoS{}, \AlVS{}), 30 modes in the orthorhombic phase $Imm2$ (\GeVS{}), and 58 modes in the tetragonal phase $P\overline{4}2_1m$ (\GaNbS{}, \GaTaSe{}). However, in each compound, only a limited subset of these IR allowed modes could be observed, especially in the low-temperature phase. The remaining modes seem to have too low optical weight to be detected.

The phonon spectra of \GaVS{} and \GeVS{} have already been discussed in detail in Ref. \cite{Reschke:2017}. For both compounds four phonon modes out of the six IR-allowed $F_2$ modes can be identified from the FIR reflectivity spectra at room temperature. In the case of \GaVS{} the two remaining $F_2$ modes have been identified by Raman spectroscopy near 130 and 190\,cm$^{-1}$. For \GaVS{}, two additional phonon modes appear below the structural phase transition at \TJT{} = 44\,K as a result of the lowering of the crystal symmetry from cubic to rhombohedral. The magnetic ordering at \TC{} = 12.7\,K has no effect on either the number of modes or the phonon parameters. In the case of \GeVS{}, at the structural transition at \TJT{} = 30.5\,K, three additional phonon modes appear, when the symmetry changes from cubic to orthorhombic. In contrast to \GaVS{}, in \GeVS{} the phonon spectrum is also affected by the magnetic transition at \TN{} = 14.6\,K \cite{Reschke:2017,Widmann:2016}. At the antiferromagnetic transition, a further small splitting of three modes and changes in the phonon parameters can be observed (green arrows in Fig. \ref{fig:FIR_R}). Since the degeneracy of the modes should already be fully lifted upon the orthorhombic transition, the additional splitting is probably due to the increased size of the unit cell in the antiferromagnetic state and spin-lattice coupling \cite{Reschke:2017}. Although there are differences in the oscillator strengths, the room-temperature eigenfrequencies of the observed phonon modes of \GaVS{} and \GeVS{} are very similar, most likely because Ga and Ge are neighbors in the periodic table with very similar masses.

In the case of \GaMoS{}, again four out of six IR-allowed phonon modes are observed in the high-temperature cubic phase, with two of them dominating, located at 305 and 370\,cm$^{-1}$. On lowering the temperature, six additional phonon modes appear at the structural transition at \TJT{} = 45\,K, all of which have very low spectral weight compared to the two strong phonon modes. For better visibility of the additional weak phonon modes, the insets in Fig. \ref{fig:FIR_R} show the enlarged low-temperature reflectivity in comparison with the reflectivity measured at 50\,K. In this compound, the phonon spectrum is not affected by the magnetic phase transition. The phonon eigenfrequencies are again very similar to the two former compounds. No phonon modes with considerable spectral weight are found below 300\,cm$^{-1}$. Surprisingly, the replacement of V by Mo does not show a strong effect on the phonon frequencies, although the masses of V and Mo differ almost by a factor of two.

\begin{figure*}
\includegraphics[width = 1\textwidth]{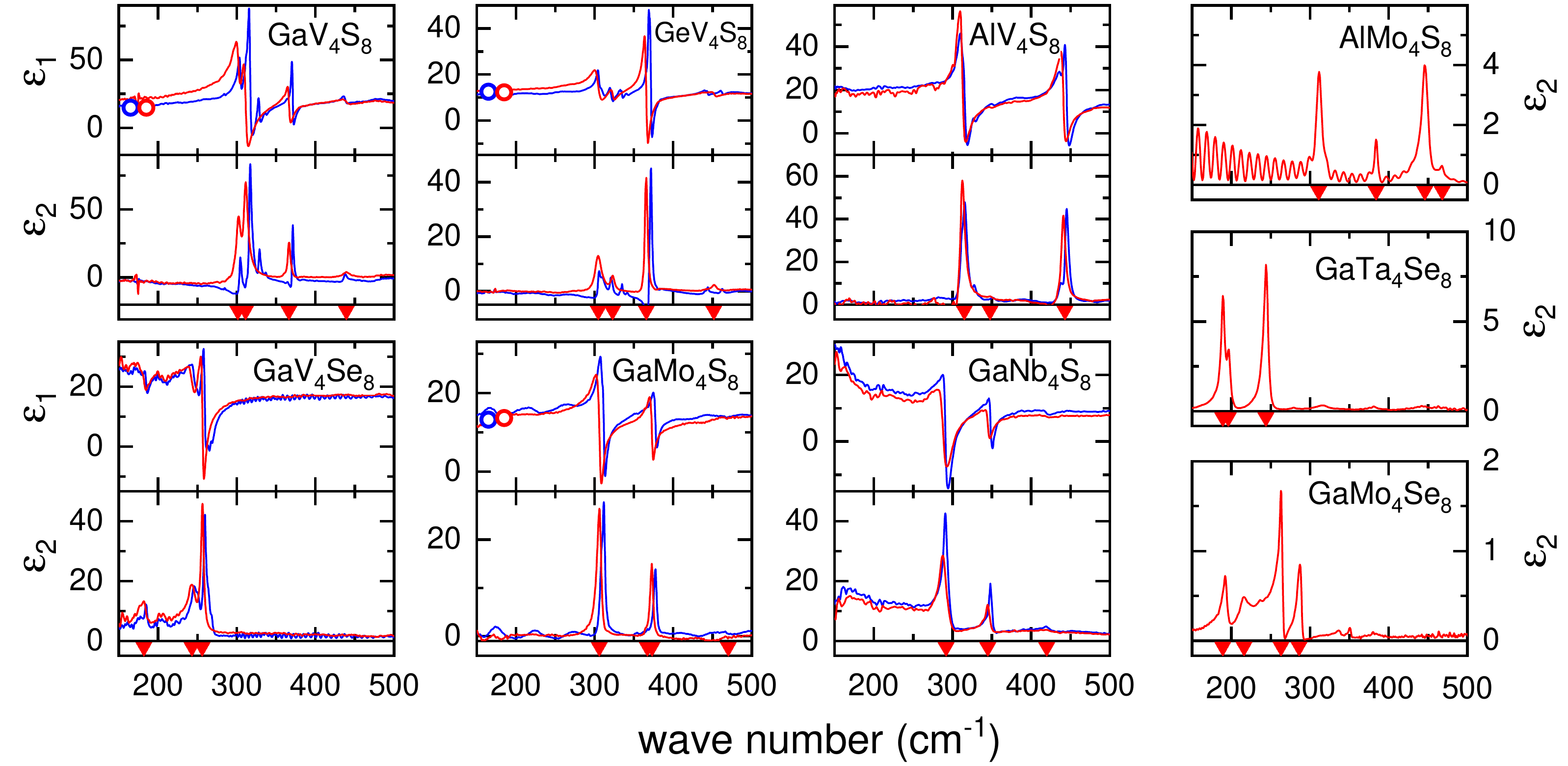}
\caption{\label{fig:epsilon} Frequency dependence of the real part ($\epsilon_1$) and the imaginary part ($\epsilon_2$) of the dielectric function in the FIR range, measured at room temperature (red curves) and at 10\,K (blue curves). Red triangles indicate the phonon eigenfrequencies at room temperature. Open circles indicate the dielectric function values obtained from THz transmission measurements at 100\,cm$^{-1}$.}
\end{figure*}

In the case of \GaNbS{} three phonon modes are identified in the room-temperature reflectivity spectra, which have relative strengths similar to those in \GaMoS{}. When comparing the room-temperature eigenfrequencies of \GaNbS{} to those of \GaMoS{}  and \GaVS{}, whose compositions differ only in the M site, the phonon modes seem to be slightly shifted towards lower frequencies. Nevertheless, the phonon frequencies seem not to be strongly influenced by changes in the M site ions. Below the magneto-structural transition, one additional phonon mode is observed.

At room temperature, three phonon modes are observed in \AlVS{}, with two stronger modes at 315\,cm$^{-1}$ and 443\,cm$^{-1}$, dominating the reflectivity spectrum. Below the structural phase transition three additional modes appear. Despite the large difference of the Ga and Al masses, the phonon eigenfrequencies of \AlVS{} are comparable to those of the other sulfide compounds. However, the relative strengths of the phonon modes in \AlVS{} and in the aforementioned Ga-based compounds are different.

From polycrystalline transmission data of \AlMoS{} four phonon modes can be identified at room temperature. The phonon frequencies and particularly the relative mode strengths show strong similarity to \AlVS{}. These results suggest that the eigenfrequencies of the phonon modes observed in lacunar spinels are not essentially related to the A sites. However, the comparison of the Ga and Al compounds implies an influence of the A-site ion on the relative strength of the phonon modes.

In contrast, when exchanging the ligand position from S to Se, it becomes apparent for the first inspection that the phonon eigenfrequencies are considerably shifted to lower frequencies. For \GaVSe{} three phonon modes could be observed at room temperature in the frequency range between 180\,cm$^{-1}$ and 270\,cm$^{-1}$. Below the symmetry-lowering structural phase transition, taking place at \TJT{} = 42 K, three additional phonon modes appear, whereas the phonon spectrum is again not affected by the magnetic ordering at \TC{} = 18 K. 

In the room-temperature transmission spectrum of \GaMoSe{} four phonon modes can be found below 300\,cm$^{-1}$, which have eigenfrequencies similar to those in \GaVSe{}. For \GaTaSe{} three phonons can be found in the same frequency range. This implies again that an exchange of the M position does not have a major effect on the phonon frequencies. Instead, the experimental results demonstrate that the phonon eigenfrequencies of the cubic high-temperature phase are mainly influenced by the choice of the ligand. These findings suggest that the observed phonon modes of the lacunar spinels under investigation are dominated by vibrations of the ligands. Indeed, when considering the difference between S and Se in terms of their masses, the ratio of the eigenfrequencies of the S and the Se compounds is very close to $\sqrt{m_{\mathrm{S}}/m_{\mathrm{Se}}} = 0.64  $, which is expected in this simple approximation. For both \GaMoSe{} and \GaTaSe{} one additional mode could be found below the structural phase transition temperature.

Figure \ref{fig:epsilon} shows the frequency-dependent dielectric function of the studied lacunar spinels at room temperature (red curves) and at 10\,K (blue curves). From reflectivity data both the real ($\epsilon_1$) and the imaginary part ($\epsilon_2$) can be directly calculated via the VDF method \cite{Kuzmenko:2005,Kuzmenko:2018}, while the transmission data, measured on polycrystalline powder samples embedded in PE, just allow an estimation of the imaginary part $\epsilon_2$ by considering an effective sample thickness. In the case of \GaMoS{}, the low-frequency upturn of the reflectivity, possibly coming from surface conductivity, was modeled by a Drude term. The dielectric function of the Drude term is given by
\begin{equation}
\epsilon(\omega)=1 -\frac{\omega_\mathrm{p}^2}{\omega^2 +i\omega\gamma},
\end{equation}
with plasma frequency $\omega_\mathrm{p}$ and damping parameter $\gamma$. At 10\,K this Drude term can be described by the parameters $\omega_\mathrm{p}= 950$\,cm$^{-1}$ and $\gamma = 235$\,cm$^{-1}$, which are only weakly temperature dependent. This Drude term was subtracted from the dielectric function data afterward in order to obtain the dielectric spectrum characteristic to the bulk. For \GaVS{}, \GeVS{}, \GaVSe{}, \GaMoS{}, \AlMoS{} and \GaNbS{} the absolute values of $\epsilon_1$ and $\epsilon_2$ are all of the same order of magnitude. The high-frequency $\epsilon_1$ values are in the range of 10 - 20 for all compounds. Furthermore, for those compounds, where THz transmission measurements were performed, the low-frequency $\epsilon_1$ values are in agreement with the THz data. This is demonstrated by the open symbols in Fig. \ref{fig:epsilon}, which depict the $\epsilon_1$ values from THz transmission at 100\,cm$^{-1}$. For \GaVS{}, \GeVS{}, \GaVSe{}, \GaMoS{}, \AlMoS{} and \GaNbS{} $\epsilon_2$ reaches values up to 25 – 50 at the resonance frequencies. As a general feature,  $\epsilon_1$ in the static limit as well as above the phonon excitations is quite high in all of the compounds, which is a consequence of the small band gap, as will be shown later. As already mentioned, the  $\epsilon_2$ spectra of \AlMoS{}, \GaTaSe{} and \GaMoSe{} shown in Fig. \ref{fig:epsilon} give a rough estimate of the absolute values. For these polycrystalline samples the values estimated are approximately one order of magnitude smaller than in the aforementioned single-crystalline samples, indicating that the absolute value of  $\epsilon_2$ is underestimated in these cases.

\begin{figure}[tb]
\includegraphics[width = 0.9\columnwidth]{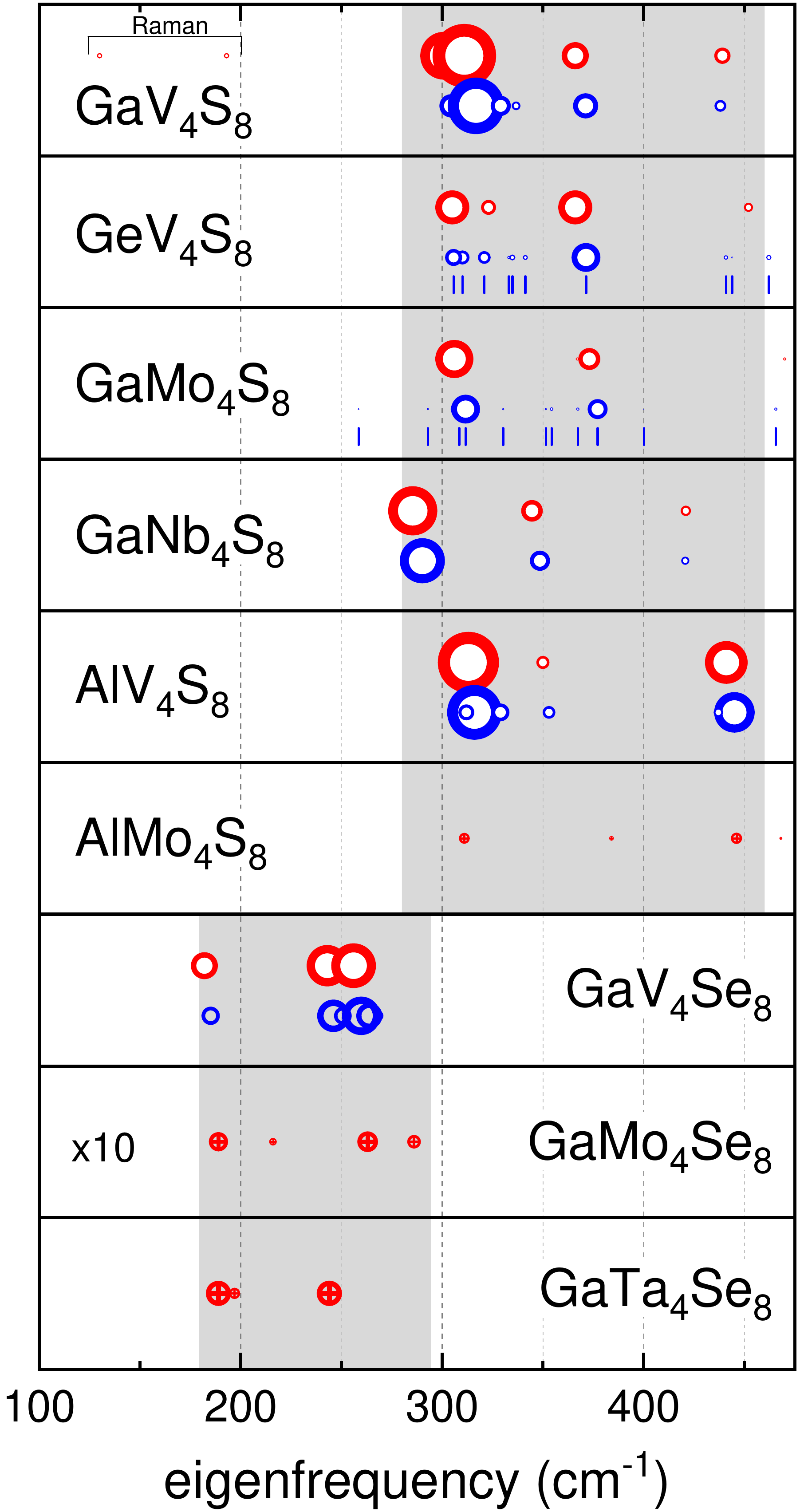}
\caption{\label{fig:phonon_summary} Summary of the eigenfrequencies of the investigated lacunar spinels at room temperature (red symbols) and at 10\,K (blue symbols). The area of the symbols represents the oscillator strengths $\Delta\epsilon$ of the phonon modes. In the case of very weak modes, the eigenfrequencies are also indicated by vertical lines.  For \GaVS{} also the two $F_2$ modes observed by Raman spectroscopy \cite{Hlinka:2016} are indicated. The shaded areas indicate the frequency ranges, within which the strongest phonon modes are located (see the main text).}
\end{figure}

Figure \ref{fig:phonon_summary} gives an overview of the comparative FIR phonon mode analysis in lacunar spinels. For each compound the eigenfrequencies at room temperature (red symbols) and 10\,K (blue symbols) are shown. The area of the symbols represents the oscillator strengths, $\Delta\epsilon$, corresponding to the individual phonon modes, which have been determined by fitting the experimental reflectivity and transmission data by Lorentz oscillators. The dielectric function of $N$ Lorentz oscillators is given by
\begin{equation}
\epsilon(\omega)=\epsilon_\infty + \sum_{i=1}^{N}\frac{\Delta\epsilon_i\omega_{0,i}^2}{\omega_{0,i}^2 -\omega^2 -i\omega\gamma_i}
\end{equation}
Here, $\omega_0$ and $\gamma$ denote the eigenfrequency and the damping of the Lorentz oscillator. $\epsilon_\infty$ is the high-frequency dielectric constant.
Again, the values of $\Delta\epsilon$, obtained from transmission measurements on polycrystalline samples, should only be treated as rough estimates. Figure \ref{fig:phonon_summary} clearly emphasizes that the phonon eigenfrequencies are primarily affected by the choice of the element at the ligand site, with Se shifting the phonon modes to lower frequencies compared to S. This systematic frequency shift can be simply interpreted by the ratio of the S and Se masses. This is indicated by the shaded areas in Fig. 5 covering the frequency ranges containing the strongest phonon modes. For the selenides this area scales with a factor of $\sqrt{m_{\mathrm{S}}/m_{\mathrm{Se}}} = 0.64  $ compared to the sulfur compounds. On the other hand the room temperature phonon eigenfrequencies do not experience major changes when replacing Ga with Ge or Al on the A sites, with the latter having a much lower mass. This scenario also seems to be very similar for the M site, because also replacing V by Mo or Nb did not result in significant changes of the room-temperature phonon frequencies. Therefore, the observed phonon modes in the lacunar spinels seem not to be related to the M site. Please note that only a subset of the IR-active modes is observed in all compounds both in the cubic and the low-temperature phases.

Comparing the relative oscillator strength of the sulfides, clear similarities are found for the Ga compounds. The strongest phonon modes appear close to 300\,cm$^{-1}$, a second relatively strong mode can be found around 350 to 380\,cm$^{-1}$ and a relatively weak mode shows up around 450\,cm$^{-1}$. Although the change from Ga to Al does not significantly change the phonon frequencies, it affects the distribution of the spectral weight on the modes. For the two Al-based compounds the modes around 300 and 450\,cm$^{-1}$ have the highest oscillator strength, while the mode in between gets relatively weak. From this we conclude that the atoms at the A sites are clearly involved in the phonon mode located between 350 - 380\,cm$^{-1}$ (depending on the material) and in the mode located at around 450\,cm$^{-1}$. In the case of the selenide compounds the two strongest modes are located at around 190\,cm$^{-1}$ and between 250 - 280\,cm$^{-1}$. Major changes in the relative oscillator strengths due to exchanging the M-site ion are not discerned. As it can be seen from Fig. \ref{fig:phonon_summary}, especially for the low-temperature phases, which do not all have the same crystal symmetry, a direct comparison of eigenfrequency and oscillator strength of every single phonon mode is not straightforward.

Concerning the number of phonons, for each of the studied lacunar spinels only a subset of the six allowed zone center $F_2$ modes, either three or four, was observed at room temperature. When only considering the ligands (at X$^1$ and X$^2$ positions) and the A site ions, the symmetry analysis results in four triply degenerate $F_2$ IR-active phonon modes, after removing one $F_2$ mode related to the acoustic modes. Therefore, the observed IR phonon modes may be ascribed to vibrations of the AX$_4$ clusters and to the ligands within the M$_4$X$_4$ cubane clusters, without discernible contribution of the M$_4$ tetrahedra. This is also in line with the systematic changes in the phonon spectra over the broad series of the studied lacunar spinel compounds: The exchange of the ligand gives rise to a clear shift of all observed modes, while the replacement of the A-site ion changes the relative intensities of two of them. In contrast, no effect of the replacement of the M-site ion could be observed.

\subsection{\label{sec:BroadbandConductivity}Broadband optical conductivity}

\begin{figure}[tb]
\includegraphics[width = 0.9\columnwidth]{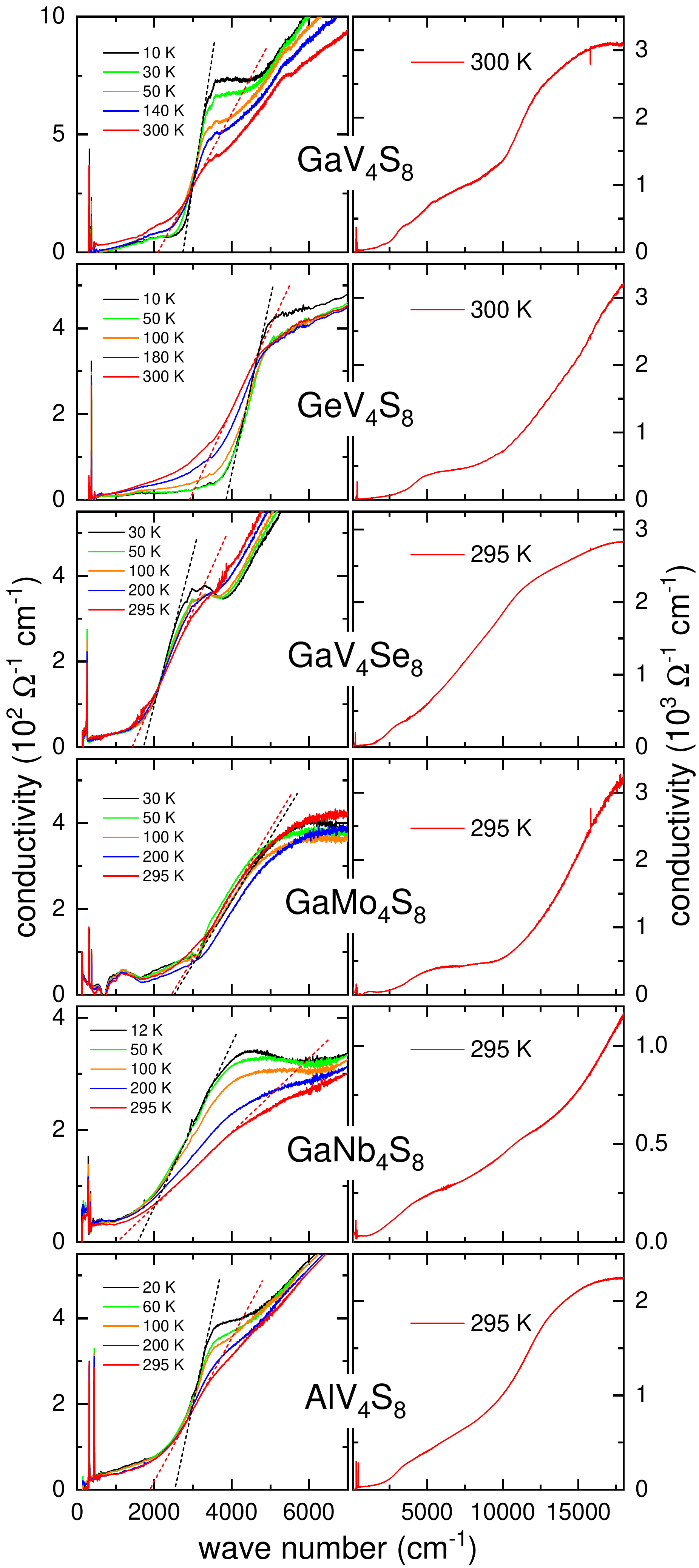}
\caption{\label{fig:conductivity} Frequency-dependent optical conductivity of \GaVS{}, \GeVS{}, \GaVSe{}, \GaMoS{}, \GaNbS{} and \AlVS{} for frequencies up to 7000\,cm$^{-1}$ for selected temperatures between 10 and 300\,K (left panels). The dashed lines indicate linear fits of the conductivity increase ascribed to the band gap. Broadband optical conductivity at room temperature up to 18000\,cm$^{-1}$ (right panels).}
\end{figure}

In order to gain insight into the electronic structure of lacunar spinels the broadband optical conductivity of the single-crystalline samples was analyzed in the frequency range below 20000\,cm$^{-1}$. Figure \ref{fig:conductivity} shows the frequency dependent optical conductivity of \GaVS{}, \GeVS{}, \GaVSe{}, \GaMoS{}, \GaNbS{} and \AlVS{} for a series of temperatures between 100 and 300\,K in the frequency range up to 7000\,cm$^{-1}$ as well as the room-temperature broadband optical conductivity up to 18000\,cm$^{-1}$. At low frequencies the optical conductivity vanishes in all compounds, clearly demonstrating the semiconductor nature of the compounds even in the room-temperature cubic state. In the frequency range between 2000 - 5000\,cm$^{-1}$ an increase in the conductivity can be observed for all of them, which represent the band edge, associated to the band gap. In \GaVS{} and \GeVS{}, the band edge shows a strong temperature dependence: For the lowest temperatures the band edge manifests itself in a sudden increase of the optical conductivity at around 3000 and 4000\,cm$^{-1}$, respectively, while the band edge gets smeared out with increasing temperature. This has been explained by orbital fluctuations in the high-temperature phase \cite{Reschke:2017} and changes in the orbital occupancy upon the orbitally driven structural transition. A similar behavior is found in \GaVSe{}, \GaNbS{} and \AlVS{}. In contrast, in the optical conductivity spectra of \GaMoS{} no significant systematic temperature dependence of the band edge is followed.

To estimate the value of the band gap, a linear extrapolation of the optical conductivities to zero has been performed in the region of the band edge, as shown by dashed lines in Fig. \ref{fig:conductivity}. The gap values $E_\mathrm{g}$ of the studied compounds at low temperature and at 300\,K are summarized in Table \ref{tab:gap}, where also a comparison with literature values determined from resistivity and optical measurements is given. For \GaVS{} this yields a band gap of 260\,meV at room temperature, which is very close to the value determined by resistivity measurements \cite{Widmann:2016a}. With decreasing temperature the band gap shifts to a considerably higher energy of 340\,meV at 10\,K. The gap values of \GeVS{} are slightly higher. At room temperature a band gap of 350\,meV is found, which is in reasonable agreement with the value of 300\,meV reported in the literature \cite{Widmann:2016}. Again the band gap experiences a strong blue-shift up to 475\,meV at 10\,K. In comparison to \GaVS{} and \GeVS{}, the energy gap of \GaVSe{} is slightly smaller and the temperature dependence of the gap value is less pronounced. Here a room-temperature gap of 175\,meV can be found, which increases up to 210\,meV at the lowest temperature. In the case of \GaMoS{} the optical conductivity data do not show a clear trend in the temperature evolution of the band gap. Both for the highest and the lowest temperature a band gap of 300\,meV can be determined. For \GaNbS{} the optical conductivity exhibits a pronounced temperature dependence. For the band gap relatively low values of 130\,meV at room temperature and 195\,meV at\,12 K could be determined. The linear extrapolation of the optical conductivity of \AlVS{} yields a gap of 230\,meV at room temperature, which is increased up to 310\,meV at the lowest temperature. These results from optical conductivity analysis confirm that lacunar spinels are narrow-gap insulators with gap energies between 200 and 330\,meV  \cite{Cario:2010}.

\begin{table}[tb]
\caption{\label{tab:gap}
Band gap $E_\mathrm{g}$ of lacunar spinels determined from the optical conductivity spectra at low temperatures and at 300 K. In comparison, values determined from resistivity measurements are also shown. Values without reference have been determined from the optical conductivity in the course of this work. All values are given in meV.}
\begin{ruledtabular}
\begin{tabular}{lccc}
 & \multicolumn{2}{c}{$E_\mathrm{g}$ optics} & $E_\mathrm{g}$ resistivity \\
 & low $T$ & 300\,K &  \\
\hline
\GaVS{}     &   340 (10\,K) &   260 &  240 \cite{Widmann:2016a}   \\
            &               &       &  330 \cite{Guiot:2013} \\
\GeVS{}     &   475 (10\,K) &   350 &  300 \cite{Widmann:2016}   \\
\AlVS{}     &   310 (20\,K) &   230 &    \\
\GaMoS{}    &   300 (30\,K) &   300 &    \\
\GaNbS{}    &   195 (12\,K) &   130 &  280 \cite{Pocha:2005}   \\
\GaVSe{}    &   210 (30\,K) &   175 &  270 \cite{Bichler:2011}   \\
\GaNbSe{}   &               &       &  190 \cite{Pocha:2005}   \\
            &               &       &  280 \cite{Guiot:2013}   \\
\GaTaSe{}   &               &   120 \cite{Guiot:2013}     &  160 \cite{Pocha:2005}  \\
            &               &       &  240 \cite{Guiot:2011}   \\
\end{tabular}
\end{ruledtabular}
\end{table}

When comparing the gap energies $E_\mathrm{g}$ for the different lacunar spinel compounds when exchanging either the M or X site, the following statements can be made: Changing the ligand site from S to Se results in a reduction of $E_\mathrm{g}$, as can be seen when comparing the optical data of \GaVS{} and \GaVSe{}. A similar reduction of the gap is observed, on the basis of resistivity measurements from \GaNbS{} \cite{Pocha:2005} to \GaNbSe{} \cite{Pocha:2005}. This tendency is likely due to the larger bandwidth of the cluster orbitals in the Se-based compounds. However, conclusions based on gap values solely based on resistivity measurements should be taken with care. As is clear from Table \ref{tab:gap}, gap values based on resistivity data are rather inaccurate, likely due to strong deviations from the Arrhenius-like temperature dependence, making the determination of $E_\mathrm{g}$ somewhat arbitrary.

Exchanging the M site from $3d$ to $4d$ elements, while leaving the other elements unchanged, also results in a reduction of the gap energy, as can be seen when comparing the low-temperature gap values for \GaVS{} with \GaNbS{} and \GaMoS{}. The same decreasing tendency of $E_\mathrm{g}$ can be found when replacing V by the $5d$ element Ta, as can be seen by comparison of \GaVSe{} and \GaTaSe{}.

\begin{figure}[tb]
\includegraphics[width = 0.85\columnwidth]{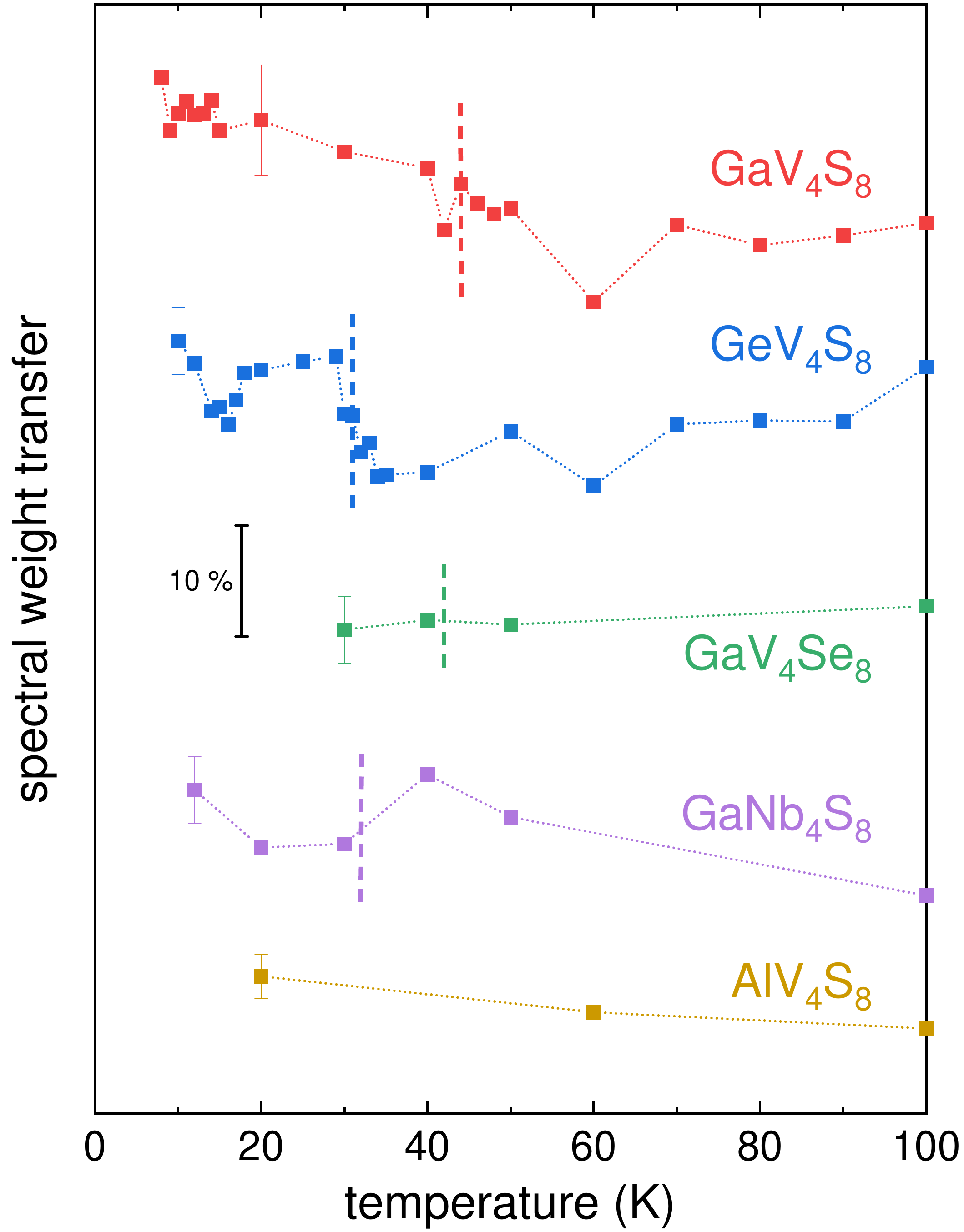}
\caption{\label{fig:spectralWeight} Temperature dependence of the spectral weight transfer relative to 100\,K determined by integration of the optical conductivities up to 6000\,cm$^{-1}$ in several compounds. The vertical dashed lines mark the corresponding structural phase transition temperatures. The curves are shifted for clarity.}
\end{figure}

In order to gain insight into the changes of the electronic structure upon the structural transition, in Fig. \ref{fig:spectralWeight} we show the temperature dependence of the spectral weight transfer relative to 100\,K, determined by integration of the optical conductivity, for five compounds. An upper integration limit of 6000\,cm$^{-1}$ has been chosen so that the frequency range of the band edge is covered in each compound. Although in \GaVS{}, \GeVS{} and \GaNbS{} weak anomalies are observed at the structural phase transition, we can conclude that the structural transition in these compounds has little effect on the band edge and, thus, the value of the gap.

\section{Conclusions}

To conclude, broadband infrared spectroscopy has been performed on a broad series of lacunar spinels. By sequentially replacing ions at the A and M sites as well as by substituting the ligand, we observe that the eigenfrequencies of the infrared active phonon modes are mainly affected by the change of the ligand: The shift of the phonons towards lower frequencies when substituting S with Se can be understood by their mass ratio. In addition, for compounds with different ions on the A sites, we observed changes in the spectral weight of the phonon modes. From this we conclude the strongest phonon modes may be ascribed to the ligand and the A-site ions of the lacunar spinel structure, with no obvious contribution from the M-site ions. We do not have a clear picture of why the vibrations of the M$_4$ tetrahedra are unseen in the infrared phonon spectra. One possible explanation is that the M$_4$ clusters are metallic, thus, the vibrations of these units are effectively screened by carriers delocalized over the M$_4$ clusters. However, this interpretation would contradict the Jahn-Teller activity of these  clusters, which is generally accepted  as the driving force of the structural transitions in lacunar spinels.  

Concerning the electronic states, we found that all lacunar spinels, investigated here with $3d$, $4d$ and $5d$ ions at the M sites, are narrow-gap semiconductors with gap values ranging from 130\,meV to 350\,meV. The band gap is clearly observed already in the high-temperature cubic state and only influenced weakly by the structural transition. Replacing S with Se reduces the gap value. A similar decrease of the gap value is found when replacing the $3d$ ion with $4d$ and $5d$ ions at the M sites. This indicates the weakening of the correlations from $3d$ to $5d$ and does not directly support the Hund-insulator picture, suggested by theoretical works \cite{Wang:2019,Lee:2019}.

\begin{acknowledgments}

We acknowledge enlightening discussions with S. Bord\'{a}cs, D. Szaller, J. Hlinka, D. Vollhardt, P. Lunkenheimer, H.-A. Krug von Nidda and A. Loidl. This research was partly funded by Deutsche Forschungsgemeinschaft DFG via the Transregional Collaborative Research Center TRR 80 “From Electronic Correlations to Functionality” (Augsburg, Munich, Stuttgart).  L. Prodan acknowledges support of Deutscher Akademischer Austauschdienst (DAAD).

\end{acknowledgments}

\providecommand{\noopsort}[1]{}\providecommand{\singleletter}[1]{#1}%


\begin{thebibliography}{65}%
\makeatletter
\providecommand \@ifxundefined [1]{%
 \@ifx{#1\undefined}
}%
\providecommand \@ifnum [1]{%
 \ifnum #1\expandafter \@firstoftwo
 \else \expandafter \@secondoftwo
 \fi
}%
\providecommand \@ifx [1]{%
 \ifx #1\expandafter \@firstoftwo
 \else \expandafter \@secondoftwo
 \fi
}%
\providecommand \natexlab [1]{#1}%
\providecommand \enquote  [1]{``#1''}%
\providecommand \bibnamefont  [1]{#1}%
\providecommand \bibfnamefont [1]{#1}%
\providecommand \citenamefont [1]{#1}%
\providecommand \href@noop [0]{\@secondoftwo}%
\providecommand \href [0]{\begingroup \@sanitize@url \@href}%
\providecommand \@href[1]{\@@startlink{#1}\@@href}%
\providecommand \@@href[1]{\endgroup#1\@@endlink}%
\providecommand \@sanitize@url [0]{\catcode `\\12\catcode `\$12\catcode
  `\&12\catcode `\#12\catcode `\^12\catcode `\_12\catcode `\%12\relax}%
\providecommand \@@startlink[1]{}%
\providecommand \@@endlink[0]{}%
\providecommand \url  [0]{\begingroup\@sanitize@url \@url }%
\providecommand \@url [1]{\endgroup\@href {#1}{\urlprefix }}%
\providecommand \urlprefix  [0]{URL }%
\providecommand \Eprint [0]{\href }%
\providecommand \doibase [0]{https://doi.org/}%
\providecommand \selectlanguage [0]{\@gobble}%
\providecommand \bibinfo  [0]{\@secondoftwo}%
\providecommand \bibfield  [0]{\@secondoftwo}%
\providecommand \translation [1]{[#1]}%
\providecommand \BibitemOpen [0]{}%
\providecommand \bibitemStop [0]{}%
\providecommand \bibitemNoStop [0]{.\EOS\space}%
\providecommand \EOS [0]{\spacefactor3000\relax}%
\providecommand \BibitemShut  [1]{\csname bibitem#1\endcsname}%
\let\auto@bib@innerbib\@empty
\bibitem [{\citenamefont {Barz}(1973)}]{Barz:1973}%
  \BibitemOpen
  \bibfield  {author} {\bibinfo {author} {\bibfnamefont {H.}~\bibnamefont
  {Barz}},\ }\href@noop {} {\bibfield  {journal} {\bibinfo  {journal} {Mat.
  Res. Bull.}\ }\textbf {\bibinfo {volume} {8}},\ \bibinfo {pages} {983}
  (\bibinfo {year} {1973})}\BibitemShut {NoStop}%
\bibitem [{\citenamefont {BenYaich}\ \emph {et~al.}(1984)\citenamefont
  {BenYaich}, \citenamefont {Jegaden}, \citenamefont {Potel}, \citenamefont
  {Sergent}, \citenamefont {Rastogi},\ and\ \citenamefont
  {Tournier}}]{BenYaich:1984}%
  \BibitemOpen
  \bibfield  {author} {\bibinfo {author} {\bibfnamefont {H.}~\bibnamefont
  {BenYaich}}, \bibinfo {author} {\bibfnamefont {J.}~\bibnamefont {Jegaden}},
  \bibinfo {author} {\bibfnamefont {M.}~\bibnamefont {Potel}}, \bibinfo
  {author} {\bibfnamefont {M.}~\bibnamefont {Sergent}}, \bibinfo {author}
  {\bibfnamefont {A.~K.}\ \bibnamefont {Rastogi}},\ and\ \bibinfo {author}
  {\bibfnamefont {R.}~\bibnamefont {Tournier}},\ }\href@noop {} {\bibfield
  {journal} {\bibinfo  {journal} {J. Less Common Met.}\ }\textbf {\bibinfo
  {volume} {102}},\ \bibinfo {pages} {9} (\bibinfo {year} {1984})}\BibitemShut
  {NoStop}%
\bibitem [{\citenamefont {Johrendt}(1998)}]{Johrendt:1998}%
  \BibitemOpen
  \bibfield  {author} {\bibinfo {author} {\bibfnamefont {D.}~\bibnamefont
  {Johrendt}},\ }\href@noop {} {\bibfield  {journal} {\bibinfo  {journal} {Z.
  anorg. Allg. Chem.}\ }\textbf {\bibinfo {volume} {624}},\ \bibinfo {pages}
  {952} (\bibinfo {year} {1998})}\BibitemShut {NoStop}%
\bibitem [{\citenamefont {Pocha}\ \emph {et~al.}(2000)\citenamefont {Pocha},
  \citenamefont {Johrendt},\ and\ \citenamefont {P{\"o}ttgen}}]{Pocha:2000}%
  \BibitemOpen
  \bibfield  {author} {\bibinfo {author} {\bibfnamefont {R.}~\bibnamefont
  {Pocha}}, \bibinfo {author} {\bibfnamefont {D.}~\bibnamefont {Johrendt}},\
  and\ \bibinfo {author} {\bibfnamefont {R.}~\bibnamefont {P{\"o}ttgen}},\
  }\href@noop {} {\bibfield  {journal} {\bibinfo  {journal} {Chem. Mater.}\
  }\textbf {\bibinfo {volume} {12}},\ \bibinfo {pages} {2882} (\bibinfo {year}
  {2000})}\BibitemShut {NoStop}%
\bibitem [{\citenamefont {K{\'{e}}zsm{\'{a}}rki}\ \emph
  {et~al.}(2015)\citenamefont {K{\'{e}}zsm{\'{a}}rki}, \citenamefont
  {Bord\'{a}cs}, \citenamefont {Milde}, \citenamefont {Neuber}, \citenamefont
  {Eng}, \citenamefont {White}, \citenamefont {Ronnow}, \citenamefont
  {Dewhurst}, \citenamefont {Mochizuki}, \citenamefont {Yanai}, \citenamefont
  {Nakamura}, \citenamefont {Ehlers}, \citenamefont {Tsurkan},\ and\
  \citenamefont {Loidl}}]{Kezsmarki:2015}%
  \BibitemOpen
  \bibfield  {author} {\bibinfo {author} {\bibfnamefont {I.}~\bibnamefont
  {K{\'{e}}zsm{\'{a}}rki}}, \bibinfo {author} {\bibfnamefont {S.}~\bibnamefont
  {Bord\'{a}cs}}, \bibinfo {author} {\bibfnamefont {P.}~\bibnamefont {Milde}},
  \bibinfo {author} {\bibfnamefont {E.}~\bibnamefont {Neuber}}, \bibinfo
  {author} {\bibfnamefont {L.~M.}\ \bibnamefont {Eng}}, \bibinfo {author}
  {\bibfnamefont {J.~S.}\ \bibnamefont {White}}, \bibinfo {author}
  {\bibfnamefont {H.~M.}\ \bibnamefont {Ronnow}}, \bibinfo {author}
  {\bibfnamefont {C.~D.}\ \bibnamefont {Dewhurst}}, \bibinfo {author}
  {\bibfnamefont {M.}~\bibnamefont {Mochizuki}}, \bibinfo {author}
  {\bibfnamefont {K.}~\bibnamefont {Yanai}}, \bibinfo {author} {\bibfnamefont
  {H.}~\bibnamefont {Nakamura}}, \bibinfo {author} {\bibfnamefont
  {D.}~\bibnamefont {Ehlers}}, \bibinfo {author} {\bibfnamefont
  {V.}~\bibnamefont {Tsurkan}},\ and\ \bibinfo {author} {\bibfnamefont
  {A.}~\bibnamefont {Loidl}},\ }\href@noop {} {\bibfield  {journal} {\bibinfo
  {journal} {Nat. Mater.}\ }\textbf {\bibinfo {volume} {14}},\ \bibinfo {pages}
  {1116} (\bibinfo {year} {2015})}\BibitemShut {NoStop}%
\bibitem [{\citenamefont {Butykai}\ \emph {et~al.}(2017)\citenamefont
  {Butykai}, \citenamefont {Bord\'{a}cs}, \citenamefont
  {K{\'{e}}zsm{\'{a}}rki}, \citenamefont {Tsurkan}, \citenamefont {Loidl},
  \citenamefont {D{\"o}ring}, \citenamefont {Neuber}, \citenamefont {Milde},
  \citenamefont {Kehr},\ and\ \citenamefont {Eng}}]{Butykai:2017}%
  \BibitemOpen
  \bibfield  {author} {\bibinfo {author} {\bibfnamefont {A.}~\bibnamefont
  {Butykai}}, \bibinfo {author} {\bibfnamefont {S.}~\bibnamefont
  {Bord\'{a}cs}}, \bibinfo {author} {\bibfnamefont {I.}~\bibnamefont
  {K{\'{e}}zsm{\'{a}}rki}}, \bibinfo {author} {\bibfnamefont {V.}~\bibnamefont
  {Tsurkan}}, \bibinfo {author} {\bibfnamefont {A.}~\bibnamefont {Loidl}},
  \bibinfo {author} {\bibfnamefont {J.}~\bibnamefont {D{\"o}ring}}, \bibinfo
  {author} {\bibfnamefont {E.}~\bibnamefont {Neuber}}, \bibinfo {author}
  {\bibfnamefont {P.}~\bibnamefont {Milde}}, \bibinfo {author} {\bibfnamefont
  {S.~C.}\ \bibnamefont {Kehr}},\ and\ \bibinfo {author} {\bibfnamefont
  {L.~M.}\ \bibnamefont {Eng}},\ }\href@noop {} {\bibfield  {journal} {\bibinfo
   {journal} {Sci. Rep.}\ }\textbf {\bibinfo {volume} {7}},\ \bibinfo {pages}
  {44663} (\bibinfo {year} {2017})}\BibitemShut {NoStop}%
\bibitem [{\citenamefont {Vandenberg}\ and\ \citenamefont
  {Brasen}(1975)}]{Vandenberg:1975}%
  \BibitemOpen
  \bibfield  {author} {\bibinfo {author} {\bibfnamefont {J.~M.}\ \bibnamefont
  {Vandenberg}}\ and\ \bibinfo {author} {\bibfnamefont {D.}~\bibnamefont
  {Brasen}},\ }\href@noop {} {\bibfield  {journal} {\bibinfo  {journal} {J.
  Solid State Chem.}\ }\textbf {\bibinfo {volume} {14}},\ \bibinfo {pages}
  {203} (\bibinfo {year} {1975})}\BibitemShut {NoStop}%
\bibitem [{\citenamefont {Harris}(1989)}]{Harris:1989}%
  \BibitemOpen
  \bibfield  {author} {\bibinfo {author} {\bibfnamefont {S.}~\bibnamefont
  {Harris}},\ }\href@noop {} {\bibfield  {journal} {\bibinfo  {journal}
  {Polyhedron}\ }\textbf {\bibinfo {volume} {8}},\ \bibinfo {pages} {2843}
  (\bibinfo {year} {1989})}\BibitemShut {NoStop}%
\bibitem [{\citenamefont {M{\"u}ller}\ \emph {et~al.}(2006)\citenamefont
  {M{\"u}ller}, \citenamefont {Kockelmann},\ and\ \citenamefont
  {Johrendt}}]{Mueller:2006}%
  \BibitemOpen
  \bibfield  {author} {\bibinfo {author} {\bibfnamefont {H.}~\bibnamefont
  {M{\"u}ller}}, \bibinfo {author} {\bibfnamefont {W.}~\bibnamefont
  {Kockelmann}},\ and\ \bibinfo {author} {\bibfnamefont {D.}~\bibnamefont
  {Johrendt}},\ }\href@noop {} {\bibfield  {journal} {\bibinfo  {journal}
  {Chem. Mater.}\ }\textbf {\bibinfo {volume} {18}},\ \bibinfo {pages} {2174}
  (\bibinfo {year} {2006})}\BibitemShut {NoStop}%
\bibitem [{\citenamefont {Fran\c{c}ois}\ \emph {et~al.}(1991)\citenamefont
  {Fran\c{c}ois}, \citenamefont {Lengauer}, \citenamefont {Yvon}, \citenamefont
  {Yaich-Aerrache}, \citenamefont {Gougeon}, \citenamefont {Potel},\ and\
  \citenamefont {Sergent}}]{Francois:1991}%
  \BibitemOpen
  \bibfield  {author} {\bibinfo {author} {\bibfnamefont {M.}~\bibnamefont
  {Fran\c{c}ois}}, \bibinfo {author} {\bibfnamefont {W.}~\bibnamefont
  {Lengauer}}, \bibinfo {author} {\bibfnamefont {K.}~\bibnamefont {Yvon}},
  \bibinfo {author} {\bibfnamefont {H.~B.}\ \bibnamefont {Yaich-Aerrache}},
  \bibinfo {author} {\bibfnamefont {P.}~\bibnamefont {Gougeon}}, \bibinfo
  {author} {\bibfnamefont {M.}~\bibnamefont {Potel}},\ and\ \bibinfo {author}
  {\bibfnamefont {M.}~\bibnamefont {Sergent}},\ }\href@noop {} {\bibfield
  {journal} {\bibinfo  {journal} {Z. Kristallogr.}\ }\textbf {\bibinfo {volume}
  {196}},\ \bibinfo {pages} {111} (\bibinfo {year} {1991})}\BibitemShut
  {NoStop}%
\bibitem [{\citenamefont {Fujima}\ \emph {et~al.}(2017)\citenamefont {Fujima},
  \citenamefont {Abe}, \citenamefont {Tokunaga},\ and\ \citenamefont
  {Arima}}]{Fujima:2017}%
  \BibitemOpen
  \bibfield  {author} {\bibinfo {author} {\bibfnamefont {Y.}~\bibnamefont
  {Fujima}}, \bibinfo {author} {\bibfnamefont {N.}~\bibnamefont {Abe}},
  \bibinfo {author} {\bibfnamefont {Y.}~\bibnamefont {Tokunaga}},\ and\
  \bibinfo {author} {\bibfnamefont {T.}~\bibnamefont {Arima}},\ }\href@noop {}
  {\bibfield  {journal} {\bibinfo  {journal} {Phys. Rev. B}\ }\textbf {\bibinfo
  {volume} {95}},\ \bibinfo {pages} {180410(R)} (\bibinfo {year}
  {2017})}\BibitemShut {NoStop}%
\bibitem [{\citenamefont {Fran\c{c}ois}\ \emph {et~al.}(1992)\citenamefont
  {Fran\c{c}ois}, \citenamefont {Alexandrov}, \citenamefont {Yvon},
  \citenamefont {Yaich-Aerrache}, \citenamefont {Gougeon}, \citenamefont
  {Potel},\ and\ \citenamefont {Sergent}}]{Francois:1992}%
  \BibitemOpen
  \bibfield  {author} {\bibinfo {author} {\bibfnamefont {M.}~\bibnamefont
  {Fran\c{c}ois}}, \bibinfo {author} {\bibfnamefont {O.~V.}\ \bibnamefont
  {Alexandrov}}, \bibinfo {author} {\bibfnamefont {K.}~\bibnamefont {Yvon}},
  \bibinfo {author} {\bibfnamefont {H.~B.}\ \bibnamefont {Yaich-Aerrache}},
  \bibinfo {author} {\bibfnamefont {P.}~\bibnamefont {Gougeon}}, \bibinfo
  {author} {\bibfnamefont {M.}~\bibnamefont {Potel}},\ and\ \bibinfo {author}
  {\bibfnamefont {M.}~\bibnamefont {Sergent}},\ }\href@noop {} {\bibfield
  {journal} {\bibinfo  {journal} {Z. Kristallogr.}\ }\textbf {\bibinfo {volume}
  {200}},\ \bibinfo {pages} {47} (\bibinfo {year} {1992})}\BibitemShut
  {NoStop}%
\bibitem [{\citenamefont {Jakob}\ \emph {et~al.}(2007)\citenamefont {Jakob},
  \citenamefont {M{\"u}ller}, \citenamefont {Johrendt}, \citenamefont
  {Altmannshofer}, \citenamefont {Scherer}, \citenamefont {Rayaprol},\ and\
  \citenamefont {P{\"o}ttgen}}]{Jakob:2007}%
  \BibitemOpen
  \bibfield  {author} {\bibinfo {author} {\bibfnamefont {S.}~\bibnamefont
  {Jakob}}, \bibinfo {author} {\bibfnamefont {H.}~\bibnamefont {M{\"u}ller}},
  \bibinfo {author} {\bibfnamefont {D.}~\bibnamefont {Johrendt}}, \bibinfo
  {author} {\bibfnamefont {S.}~\bibnamefont {Altmannshofer}}, \bibinfo {author}
  {\bibfnamefont {W.}~\bibnamefont {Scherer}}, \bibinfo {author} {\bibfnamefont
  {S.}~\bibnamefont {Rayaprol}},\ and\ \bibinfo {author} {\bibfnamefont
  {R.}~\bibnamefont {P{\"o}ttgen}},\ }\href@noop {} {\bibfield  {journal}
  {\bibinfo  {journal} {J. Mat. Chem.}\ }\textbf {\bibinfo {volume} {17}},\
  \bibinfo {pages} {3833} (\bibinfo {year} {2007})}\BibitemShut {NoStop}%
\bibitem [{\citenamefont {Jakob}(2007)}]{Jakob:2007a}%
  \BibitemOpen
  \bibfield  {author} {\bibinfo {author} {\bibfnamefont {S.}~\bibnamefont
  {Jakob}},\ }\href@noop {} {\bibinfo {type} {{Ph.D.} thesis}},\ \bibinfo
  {school} {Ludwig-Maximilians-Universit{\"a}t M{\"u}nchen} (\bibinfo {year}
  {2007})\BibitemShut {NoStop}%
\bibitem [{\citenamefont {Bichler}\ \emph {et~al.}(2008)\citenamefont
  {Bichler}, \citenamefont {Zinth}, \citenamefont {Johrendt}, \citenamefont
  {Heyer}, \citenamefont {Forthaus}, \citenamefont {Lorenz},\ and\
  \citenamefont {Abd-Elmeguid}}]{Bichler:2008}%
  \BibitemOpen
  \bibfield  {author} {\bibinfo {author} {\bibfnamefont {D.}~\bibnamefont
  {Bichler}}, \bibinfo {author} {\bibfnamefont {V.}~\bibnamefont {Zinth}},
  \bibinfo {author} {\bibfnamefont {D.}~\bibnamefont {Johrendt}}, \bibinfo
  {author} {\bibfnamefont {O.}~\bibnamefont {Heyer}}, \bibinfo {author}
  {\bibfnamefont {M.~K.}\ \bibnamefont {Forthaus}}, \bibinfo {author}
  {\bibfnamefont {T.}~\bibnamefont {Lorenz}},\ and\ \bibinfo {author}
  {\bibfnamefont {M.~M.}\ \bibnamefont {Abd-Elmeguid}},\ }\href@noop {}
  {\bibfield  {journal} {\bibinfo  {journal} {Phys. Rev. B}\ }\textbf {\bibinfo
  {volume} {77}},\ \bibinfo {pages} {212102} (\bibinfo {year}
  {2008})}\BibitemShut {NoStop}%
\bibitem [{\citenamefont {Bord\'{a}cs}\ \emph {et~al.}(2017)\citenamefont
  {Bord\'{a}cs}, \citenamefont {Butykai}, \citenamefont {Szigeti},
  \citenamefont {White}, \citenamefont {Cubitt}, \citenamefont {Leonov},
  \citenamefont {Widmann}, \citenamefont {Ehlers}, \citenamefont {{Krug von
  Nidda}}, \citenamefont {Tsurkan}, \citenamefont {Loidl},\ and\ \citenamefont
  {K{\'{e}}zsm{\'{a}}rki}}]{Bordacs:2017}%
  \BibitemOpen
  \bibfield  {author} {\bibinfo {author} {\bibfnamefont {S.}~\bibnamefont
  {Bord\'{a}cs}}, \bibinfo {author} {\bibfnamefont {A.}~\bibnamefont
  {Butykai}}, \bibinfo {author} {\bibfnamefont {B.~G.}\ \bibnamefont
  {Szigeti}}, \bibinfo {author} {\bibfnamefont {J.~S.}\ \bibnamefont {White}},
  \bibinfo {author} {\bibfnamefont {R.}~\bibnamefont {Cubitt}}, \bibinfo
  {author} {\bibfnamefont {A.~O.}\ \bibnamefont {Leonov}}, \bibinfo {author}
  {\bibfnamefont {S.}~\bibnamefont {Widmann}}, \bibinfo {author} {\bibfnamefont
  {D.}~\bibnamefont {Ehlers}}, \bibinfo {author} {\bibfnamefont {H.-A.}\
  \bibnamefont {{Krug von Nidda}}}, \bibinfo {author} {\bibfnamefont
  {V.}~\bibnamefont {Tsurkan}}, \bibinfo {author} {\bibfnamefont
  {A.}~\bibnamefont {Loidl}},\ and\ \bibinfo {author} {\bibfnamefont
  {I.}~\bibnamefont {K{\'{e}}zsm{\'{a}}rki}},\ }\href@noop {} {\bibfield
  {journal} {\bibinfo  {journal} {Sci. Rep.}\ }\textbf {\bibinfo {volume}
  {7}},\ \bibinfo {pages} {7584} (\bibinfo {year} {2017})}\BibitemShut
  {NoStop}%
\bibitem [{\citenamefont {{A. Butykai and D. Szaller and L. F. Kiss and L.
  Balogh and M. Garst and L. DeBeer-Schmitt and T. Waki and Y. Tabata and H.
  Nakamura and I. K{\'{e}}zsm{\'{a}}rki and S.
  Bord\'{a}cs}}(2019)}]{Butykai:2019}%
  \BibitemOpen
  \bibfield  {author} {\bibinfo {author} {\bibnamefont {{A. Butykai and D.
  Szaller and L. F. Kiss and L. Balogh and M. Garst and L. DeBeer-Schmitt and
  T. Waki and Y. Tabata and H. Nakamura and I. K{\'{e}}zsm{\'{a}}rki and S.
  Bord\'{a}cs}}},\ }\href@noop {} {\bibfield  {journal} {\bibinfo  {journal}
  {arXiv:1910.11523v1}\ } (\bibinfo {year} {2019})}\BibitemShut {NoStop}%
\bibitem [{\citenamefont {Wang}\ \emph {et~al.}(2015)\citenamefont {Wang},
  \citenamefont {Ruff}, \citenamefont {Schmidt}, \citenamefont {Tsurkan},
  \citenamefont {K{\'{e}}zsm{\'{a}}rki}, \citenamefont {Lunkenheimer},\ and\
  \citenamefont {Loidl}}]{Wang:2015}%
  \BibitemOpen
  \bibfield  {author} {\bibinfo {author} {\bibfnamefont {Z.}~\bibnamefont
  {Wang}}, \bibinfo {author} {\bibfnamefont {E.}~\bibnamefont {Ruff}}, \bibinfo
  {author} {\bibfnamefont {M.}~\bibnamefont {Schmidt}}, \bibinfo {author}
  {\bibfnamefont {V.}~\bibnamefont {Tsurkan}}, \bibinfo {author} {\bibfnamefont
  {I.}~\bibnamefont {K{\'{e}}zsm{\'{a}}rki}}, \bibinfo {author} {\bibfnamefont
  {P.}~\bibnamefont {Lunkenheimer}},\ and\ \bibinfo {author} {\bibfnamefont
  {A.}~\bibnamefont {Loidl}},\ }\href@noop {} {\bibfield  {journal} {\bibinfo
  {journal} {Phys. Rev. Lett.}\ }\textbf {\bibinfo {volume} {115}},\ \bibinfo
  {pages} {207601} (\bibinfo {year} {2015})}\BibitemShut {NoStop}%
\bibitem [{\citenamefont {Ruff}\ \emph {et~al.}(2017)\citenamefont {Ruff},
  \citenamefont {Butykai}, \citenamefont {Geirhos}, \citenamefont {Widmann},
  \citenamefont {Tsurkan}, \citenamefont {Stefanet}, \citenamefont
  {K{\'{e}}zsm{\'{a}}rki}, \citenamefont {Loidl},\ and\ \citenamefont
  {Lunkenheimer}}]{Ruff:2017}%
  \BibitemOpen
  \bibfield  {author} {\bibinfo {author} {\bibfnamefont {E.}~\bibnamefont
  {Ruff}}, \bibinfo {author} {\bibfnamefont {A.}~\bibnamefont {Butykai}},
  \bibinfo {author} {\bibfnamefont {K.}~\bibnamefont {Geirhos}}, \bibinfo
  {author} {\bibfnamefont {S.}~\bibnamefont {Widmann}}, \bibinfo {author}
  {\bibfnamefont {V.}~\bibnamefont {Tsurkan}}, \bibinfo {author} {\bibfnamefont
  {E.}~\bibnamefont {Stefanet}}, \bibinfo {author} {\bibfnamefont
  {I.}~\bibnamefont {K{\'{e}}zsm{\'{a}}rki}}, \bibinfo {author} {\bibfnamefont
  {A.}~\bibnamefont {Loidl}},\ and\ \bibinfo {author} {\bibfnamefont
  {P.}~\bibnamefont {Lunkenheimer}},\ }\href@noop {} {\bibfield  {journal}
  {\bibinfo  {journal} {Phys. Rev. B}\ }\textbf {\bibinfo {volume} {96}},\
  \bibinfo {pages} {165119} (\bibinfo {year} {2017})}\BibitemShut {NoStop}%
\bibitem [{\citenamefont {Geirhos}\ \emph {et~al.}(2018)\citenamefont
  {Geirhos}, \citenamefont {Krohns}, \citenamefont {Nakamura}, \citenamefont
  {Waki}, \citenamefont {Tabata}, \citenamefont {K{\'{e}}zsm{\'{a}}rki},\ and\
  \citenamefont {Lunkenheimer}}]{Geirhos:2018}%
  \BibitemOpen
  \bibfield  {author} {\bibinfo {author} {\bibfnamefont {K.}~\bibnamefont
  {Geirhos}}, \bibinfo {author} {\bibfnamefont {S.}~\bibnamefont {Krohns}},
  \bibinfo {author} {\bibfnamefont {H.}~\bibnamefont {Nakamura}}, \bibinfo
  {author} {\bibfnamefont {T.}~\bibnamefont {Waki}}, \bibinfo {author}
  {\bibfnamefont {Y.}~\bibnamefont {Tabata}}, \bibinfo {author} {\bibfnamefont
  {I.}~\bibnamefont {K{\'{e}}zsm{\'{a}}rki}},\ and\ \bibinfo {author}
  {\bibfnamefont {P.}~\bibnamefont {Lunkenheimer}},\ }\href@noop {} {\bibfield
  {journal} {\bibinfo  {journal} {Phys. Rev. B}\ }\textbf {\bibinfo {volume}
  {98}},\ \bibinfo {pages} {224306} (\bibinfo {year} {2018})}\BibitemShut
  {NoStop}%
\bibitem [{\citenamefont {Abd-Elmeguid}\ \emph {et~al.}(2004)\citenamefont
  {Abd-Elmeguid}, \citenamefont {Ni}, \citenamefont {Khomskii}, \citenamefont
  {Pocha}, \citenamefont {Johrendt}, \citenamefont {Wang},\ and\ \citenamefont
  {Syassen}}]{Abd:2004}%
  \BibitemOpen
  \bibfield  {author} {\bibinfo {author} {\bibfnamefont {M.~M.}\ \bibnamefont
  {Abd-Elmeguid}}, \bibinfo {author} {\bibfnamefont {B.}~\bibnamefont {Ni}},
  \bibinfo {author} {\bibfnamefont {D.~I.}\ \bibnamefont {Khomskii}}, \bibinfo
  {author} {\bibfnamefont {R.}~\bibnamefont {Pocha}}, \bibinfo {author}
  {\bibfnamefont {D.}~\bibnamefont {Johrendt}}, \bibinfo {author}
  {\bibfnamefont {X.}~\bibnamefont {Wang}},\ and\ \bibinfo {author}
  {\bibfnamefont {K.}~\bibnamefont {Syassen}},\ }\href@noop {} {\bibfield
  {journal} {\bibinfo  {journal} {Phys. Rev. Lett.}\ }\textbf {\bibinfo
  {volume} {93}},\ \bibinfo {pages} {126403} (\bibinfo {year}
  {2004})}\BibitemShut {NoStop}%
\bibitem [{\citenamefont {Pocha}\ \emph {et~al.}(2005)\citenamefont {Pocha},
  \citenamefont {Johrendt}, \citenamefont {Ni},\ and\ \citenamefont
  {Abd-Elmeguid}}]{Pocha:2005}%
  \BibitemOpen
  \bibfield  {author} {\bibinfo {author} {\bibfnamefont {R.}~\bibnamefont
  {Pocha}}, \bibinfo {author} {\bibfnamefont {D.}~\bibnamefont {Johrendt}},
  \bibinfo {author} {\bibfnamefont {B.}~\bibnamefont {Ni}},\ and\ \bibinfo
  {author} {\bibfnamefont {M.~M.}\ \bibnamefont {Abd-Elmeguid}},\ }\href@noop
  {} {\bibfield  {journal} {\bibinfo  {journal} {J. Am. Chem. Soc.}\ }\textbf
  {\bibinfo {volume} {127}},\ \bibinfo {pages} {8732} (\bibinfo {year}
  {2005})}\BibitemShut {NoStop}%
\bibitem [{\citenamefont {{Ta Phuoc}}\ \emph {et~al.}(2013)\citenamefont {{Ta
  Phuoc}}, \citenamefont {Vaju}, \citenamefont {Corraze}, \citenamefont
  {Sopracase}, \citenamefont {Perucchi}, \citenamefont {Marini}, \citenamefont
  {Postorino}, \citenamefont {Chligui}, \citenamefont {Lupi}, \citenamefont
  {Janod},\ and\ \citenamefont {Cario}}]{Phuoc:2013}%
  \BibitemOpen
  \bibfield  {author} {\bibinfo {author} {\bibfnamefont {V.}~\bibnamefont {{Ta
  Phuoc}}}, \bibinfo {author} {\bibfnamefont {C.}~\bibnamefont {Vaju}},
  \bibinfo {author} {\bibfnamefont {B.}~\bibnamefont {Corraze}}, \bibinfo
  {author} {\bibfnamefont {R.}~\bibnamefont {Sopracase}}, \bibinfo {author}
  {\bibfnamefont {A.}~\bibnamefont {Perucchi}}, \bibinfo {author}
  {\bibfnamefont {C.}~\bibnamefont {Marini}}, \bibinfo {author} {\bibfnamefont
  {P.}~\bibnamefont {Postorino}}, \bibinfo {author} {\bibfnamefont
  {M.}~\bibnamefont {Chligui}}, \bibinfo {author} {\bibfnamefont
  {S.}~\bibnamefont {Lupi}}, \bibinfo {author} {\bibfnamefont {E.}~\bibnamefont
  {Janod}},\ and\ \bibinfo {author} {\bibfnamefont {L.}~\bibnamefont {Cario}},\
  }\href@noop {} {\bibfield  {journal} {\bibinfo  {journal} {Phys. Rev. Lett.}\
  }\textbf {\bibinfo {volume} {110}},\ \bibinfo {pages} {037401} (\bibinfo
  {year} {2013})}\BibitemShut {NoStop}%
\bibitem [{\citenamefont {Camjayi}\ \emph {et~al.}(2014)\citenamefont
  {Camjayi}, \citenamefont {Acha}, \citenamefont {Weht}, \citenamefont
  {Rodriguez}, \citenamefont {Corraze}, \citenamefont {Janod}, \citenamefont
  {Cario},\ and\ \citenamefont {Rozenberg}}]{Camjayi:2014}%
  \BibitemOpen
  \bibfield  {author} {\bibinfo {author} {\bibfnamefont {A.}~\bibnamefont
  {Camjayi}}, \bibinfo {author} {\bibfnamefont {C.}~\bibnamefont {Acha}},
  \bibinfo {author} {\bibfnamefont {R.}~\bibnamefont {Weht}}, \bibinfo {author}
  {\bibfnamefont {M.~G.}\ \bibnamefont {Rodriguez}}, \bibinfo {author}
  {\bibfnamefont {B.}~\bibnamefont {Corraze}}, \bibinfo {author} {\bibfnamefont
  {E.}~\bibnamefont {Janod}}, \bibinfo {author} {\bibfnamefont
  {L.}~\bibnamefont {Cario}},\ and\ \bibinfo {author} {\bibfnamefont {M.~J.}\
  \bibnamefont {Rozenberg}},\ }\href@noop {} {\bibfield  {journal} {\bibinfo
  {journal} {Phys. Rev. Lett.}\ }\textbf {\bibinfo {volume} {113}},\ \bibinfo
  {pages} {086404} (\bibinfo {year} {2014})}\BibitemShut {NoStop}%
\bibitem [{\citenamefont {Cario}\ \emph {et~al.}(2010)\citenamefont {Cario},
  \citenamefont {C.Vaju}, \citenamefont {Corraze}, \citenamefont {Guiot},\ and\
  \citenamefont {Janod}}]{Cario:2010}%
  \BibitemOpen
  \bibfield  {author} {\bibinfo {author} {\bibfnamefont {L.}~\bibnamefont
  {Cario}}, \bibinfo {author} {\bibnamefont {C.Vaju}}, \bibinfo {author}
  {\bibfnamefont {B.}~\bibnamefont {Corraze}}, \bibinfo {author} {\bibfnamefont
  {V.}~\bibnamefont {Guiot}},\ and\ \bibinfo {author} {\bibfnamefont
  {E.}~\bibnamefont {Janod}},\ }\href@noop {} {\bibfield  {journal} {\bibinfo
  {journal} {Adv. Mater.}\ }\textbf {\bibinfo {volume} {22}},\ \bibinfo {pages}
  {5193} (\bibinfo {year} {2010})}\BibitemShut {NoStop}%
\bibitem [{\citenamefont {Dubost}\ \emph {et~al.}(2013)\citenamefont {Dubost},
  \citenamefont {Cren}, \citenamefont {Vaju}, \citenamefont {Cario},
  \citenamefont {Corraze}, \citenamefont {Janod}, \citenamefont
  {Debontridder},\ and\ \citenamefont {Roditchev}}]{Dubost:2013}%
  \BibitemOpen
  \bibfield  {author} {\bibinfo {author} {\bibfnamefont {V.}~\bibnamefont
  {Dubost}}, \bibinfo {author} {\bibfnamefont {T.}~\bibnamefont {Cren}},
  \bibinfo {author} {\bibfnamefont {C.}~\bibnamefont {Vaju}}, \bibinfo {author}
  {\bibfnamefont {L.}~\bibnamefont {Cario}}, \bibinfo {author} {\bibfnamefont
  {B.}~\bibnamefont {Corraze}}, \bibinfo {author} {\bibfnamefont
  {E.}~\bibnamefont {Janod}}, \bibinfo {author} {\bibfnamefont
  {F.}~\bibnamefont {Debontridder}},\ and\ \bibinfo {author} {\bibfnamefont
  {D.}~\bibnamefont {Roditchev}},\ }\href@noop {} {\bibfield  {journal}
  {\bibinfo  {journal} {Nano Lett.}\ }\textbf {\bibinfo {volume} {13}},\
  \bibinfo {pages} {3648} (\bibinfo {year} {2013})}\BibitemShut {NoStop}%
\bibitem [{\citenamefont {Stoliar}\ \emph {et~al.}(2013)\citenamefont
  {Stoliar}, \citenamefont {Cario}, \citenamefont {Janod}, \citenamefont
  {Corraze}, \citenamefont {Guillot-Deudon}, \citenamefont {Salmon-Bourmand},
  \citenamefont {Guiot}, \citenamefont {Tranchant},\ and\ \citenamefont
  {Rozenberg}}]{Stoliar:2013}%
  \BibitemOpen
  \bibfield  {author} {\bibinfo {author} {\bibfnamefont {P.}~\bibnamefont
  {Stoliar}}, \bibinfo {author} {\bibfnamefont {L.}~\bibnamefont {Cario}},
  \bibinfo {author} {\bibfnamefont {E.}~\bibnamefont {Janod}}, \bibinfo
  {author} {\bibfnamefont {B.}~\bibnamefont {Corraze}}, \bibinfo {author}
  {\bibfnamefont {C.}~\bibnamefont {Guillot-Deudon}}, \bibinfo {author}
  {\bibfnamefont {S.}~\bibnamefont {Salmon-Bourmand}}, \bibinfo {author}
  {\bibfnamefont {V.}~\bibnamefont {Guiot}}, \bibinfo {author} {\bibfnamefont
  {J.}~\bibnamefont {Tranchant}},\ and\ \bibinfo {author} {\bibfnamefont
  {M.}~\bibnamefont {Rozenberg}},\ }\href@noop {} {\bibfield  {journal}
  {\bibinfo  {journal} {Adv. Mater.}\ }\textbf {\bibinfo {volume} {25}},\
  \bibinfo {pages} {3222} (\bibinfo {year} {2013})}\BibitemShut {NoStop}%
\bibitem [{\citenamefont {Dorolti}\ \emph {et~al.}(2010)\citenamefont
  {Dorolti}, \citenamefont {Cario}, \citenamefont {Corraze}, \citenamefont
  {Janod}, \citenamefont {Vaju}, \citenamefont {Koo}, \citenamefont {Kan},\
  and\ \citenamefont {Whangbo}}]{Dorolti:2010}%
  \BibitemOpen
  \bibfield  {author} {\bibinfo {author} {\bibfnamefont {E.}~\bibnamefont
  {Dorolti}}, \bibinfo {author} {\bibfnamefont {L.}~\bibnamefont {Cario}},
  \bibinfo {author} {\bibfnamefont {B.}~\bibnamefont {Corraze}}, \bibinfo
  {author} {\bibfnamefont {E.}~\bibnamefont {Janod}}, \bibinfo {author}
  {\bibfnamefont {C.}~\bibnamefont {Vaju}}, \bibinfo {author} {\bibfnamefont
  {H.-J.}\ \bibnamefont {Koo}}, \bibinfo {author} {\bibfnamefont
  {E.}~\bibnamefont {Kan}},\ and\ \bibinfo {author} {\bibfnamefont {M.-H.}\
  \bibnamefont {Whangbo}},\ }\href@noop {} {\bibfield  {journal} {\bibinfo
  {journal} {J. Am. Chem. Soc.}\ }\textbf {\bibinfo {volume} {132}},\ \bibinfo
  {pages} {5704} (\bibinfo {year} {2010})}\BibitemShut {NoStop}%
\bibitem [{\citenamefont {Kim}\ \emph {et~al.}(2014)\citenamefont {Kim},
  \citenamefont {Im}, \citenamefont {Han},\ and\ \citenamefont
  {Jin}}]{Kim:2014}%
  \BibitemOpen
  \bibfield  {author} {\bibinfo {author} {\bibfnamefont {H.-S.}\ \bibnamefont
  {Kim}}, \bibinfo {author} {\bibfnamefont {J.}~\bibnamefont {Im}}, \bibinfo
  {author} {\bibfnamefont {M.~J.}\ \bibnamefont {Han}},\ and\ \bibinfo {author}
  {\bibfnamefont {H.}~\bibnamefont {Jin}},\ }\href@noop {} {\bibfield
  {journal} {\bibinfo  {journal} {Nat. Commun.}\ }\textbf {\bibinfo {volume}
  {5}},\ \bibinfo {pages} {3988} (\bibinfo {year} {2014})}\BibitemShut
  {NoStop}%
\bibitem [{\citenamefont {Singh}\ \emph {et~al.}(2014)\citenamefont {Singh},
  \citenamefont {Simon}, \citenamefont {Cannuccia}, \citenamefont {Lepetit},
  \citenamefont {Corraze}, \citenamefont {Janod},\ and\ \citenamefont
  {Cario}}]{Singh:2014}%
  \BibitemOpen
  \bibfield  {author} {\bibinfo {author} {\bibfnamefont {K.}~\bibnamefont
  {Singh}}, \bibinfo {author} {\bibfnamefont {C.}~\bibnamefont {Simon}},
  \bibinfo {author} {\bibfnamefont {E.}~\bibnamefont {Cannuccia}}, \bibinfo
  {author} {\bibfnamefont {M.-B.}\ \bibnamefont {Lepetit}}, \bibinfo {author}
  {\bibfnamefont {B.}~\bibnamefont {Corraze}}, \bibinfo {author} {\bibfnamefont
  {E.}~\bibnamefont {Janod}},\ and\ \bibinfo {author} {\bibfnamefont
  {L.}~\bibnamefont {Cario}},\ }\href@noop {} {\bibfield  {journal} {\bibinfo
  {journal} {Phys. Rev. Lett.}\ }\textbf {\bibinfo {volume} {113}},\ \bibinfo
  {pages} {137602} (\bibinfo {year} {2014})}\BibitemShut {NoStop}%
\bibitem [{\citenamefont {Shanthi}\ and\ \citenamefont
  {Sarma}(1999)}]{Shanthi:1999}%
  \BibitemOpen
  \bibfield  {author} {\bibinfo {author} {\bibfnamefont {N.}~\bibnamefont
  {Shanthi}}\ and\ \bibinfo {author} {\bibfnamefont {D.~D.}\ \bibnamefont
  {Sarma}},\ }\href@noop {} {\bibfield  {journal} {\bibinfo  {journal} {J. Sol.
  State Chem.}\ }\textbf {\bibinfo {volume} {148}},\ \bibinfo {pages} {143}
  (\bibinfo {year} {1999})}\BibitemShut {NoStop}%
\bibitem [{\citenamefont {Camjayi}\ \emph {et~al.}(2012)\citenamefont
  {Camjayi}, \citenamefont {Weht},\ and\ \citenamefont
  {Rozenberg}}]{Camjayi:2012}%
  \BibitemOpen
  \bibfield  {author} {\bibinfo {author} {\bibfnamefont {A.}~\bibnamefont
  {Camjayi}}, \bibinfo {author} {\bibfnamefont {R.}~\bibnamefont {Weht}},\ and\
  \bibinfo {author} {\bibfnamefont {M.~J.}\ \bibnamefont {Rozenberg}},\
  }\href@noop {} {\bibfield  {journal} {\bibinfo  {journal} {EPL}\ }\textbf
  {\bibinfo {volume} {100}},\ \bibinfo {pages} {57004} (\bibinfo {year}
  {2012})}\BibitemShut {NoStop}%
\bibitem [{\citenamefont {Cannuccia}\ \emph {et~al.}(2017)\citenamefont
  {Cannuccia}, \citenamefont {Phuoc}, \citenamefont {Bri\`{e}re}, \citenamefont
  {Cario}, \citenamefont {Janod}, \citenamefont {Corraze},\ and\ \citenamefont
  {Lepetit}}]{Cannuccia:2017}%
  \BibitemOpen
  \bibfield  {author} {\bibinfo {author} {\bibfnamefont {E.}~\bibnamefont
  {Cannuccia}}, \bibinfo {author} {\bibfnamefont {V.~T.}\ \bibnamefont
  {Phuoc}}, \bibinfo {author} {\bibfnamefont {B.}~\bibnamefont {Bri\`{e}re}},
  \bibinfo {author} {\bibfnamefont {L.}~\bibnamefont {Cario}}, \bibinfo
  {author} {\bibfnamefont {E.}~\bibnamefont {Janod}}, \bibinfo {author}
  {\bibfnamefont {B.}~\bibnamefont {Corraze}},\ and\ \bibinfo {author}
  {\bibfnamefont {M.~B.}\ \bibnamefont {Lepetit}},\ }\href@noop {} {\bibfield
  {journal} {\bibinfo  {journal} {J. Phys. Chem. C}\ }\textbf {\bibinfo
  {volume} {121}},\ \bibinfo {pages} {3522} (\bibinfo {year}
  {2017})}\BibitemShut {NoStop}%
\bibitem [{\citenamefont {Kim}\ \emph {et~al.}(2018)\citenamefont {Kim},
  \citenamefont {Haule},\ and\ \citenamefont {Vanderbilt}}]{Kim:2018}%
  \BibitemOpen
  \bibfield  {author} {\bibinfo {author} {\bibfnamefont {H.-S.}\ \bibnamefont
  {Kim}}, \bibinfo {author} {\bibfnamefont {K.}~\bibnamefont {Haule}},\ and\
  \bibinfo {author} {\bibfnamefont {D.}~\bibnamefont {Vanderbilt}},\
  }\href@noop {} {\bibfield  {journal} {\bibinfo  {journal}
  {arXiv:1810.09495v1}\ } (\bibinfo {year} {2018})}\BibitemShut {NoStop}%
\bibitem [{\citenamefont {Wang}\ \emph {et~al.}(2019)\citenamefont {Wang},
  \citenamefont {Puggioni},\ and\ \citenamefont {Rondinelli}}]{Wang:2019}%
  \BibitemOpen
  \bibfield  {author} {\bibinfo {author} {\bibfnamefont {Y.}~\bibnamefont
  {Wang}}, \bibinfo {author} {\bibfnamefont {D.}~\bibnamefont {Puggioni}},\
  and\ \bibinfo {author} {\bibfnamefont {J.~M.}\ \bibnamefont {Rondinelli}},\
  }\href@noop {} {\bibfield  {journal} {\bibinfo  {journal} {Phys. Rev. B}\
  }\textbf {\bibinfo {volume} {100}},\ \bibinfo {pages} {115149} (\bibinfo
  {year} {2019})}\BibitemShut {NoStop}%
\bibitem [{\citenamefont {Sieberer}\ \emph {et~al.}(2007)\citenamefont
  {Sieberer}, \citenamefont {Turnovszky}, \citenamefont {Redinger},\ and\
  \citenamefont {Mohn}}]{Sieberer:2007}%
  \BibitemOpen
  \bibfield  {author} {\bibinfo {author} {\bibfnamefont {M.}~\bibnamefont
  {Sieberer}}, \bibinfo {author} {\bibfnamefont {S.}~\bibnamefont
  {Turnovszky}}, \bibinfo {author} {\bibfnamefont {J.}~\bibnamefont
  {Redinger}},\ and\ \bibinfo {author} {\bibfnamefont {P.}~\bibnamefont
  {Mohn}},\ }\href@noop {} {\bibfield  {journal} {\bibinfo  {journal} {Phys.
  Rev. B}\ }\textbf {\bibinfo {volume} {76}},\ \bibinfo {pages} {214106}
  (\bibinfo {year} {2007})}\BibitemShut {NoStop}%
\bibitem [{\citenamefont {Waki}\ \emph {et~al.}(2010)\citenamefont {Waki},
  \citenamefont {Kajinami}, \citenamefont {Tabata}, \citenamefont {Nakamura},
  \citenamefont {Yoshida}, \citenamefont {Takigawa},\ and\ \citenamefont
  {Watanabe}}]{Waki:2010}%
  \BibitemOpen
  \bibfield  {author} {\bibinfo {author} {\bibfnamefont {T.}~\bibnamefont
  {Waki}}, \bibinfo {author} {\bibfnamefont {Y.}~\bibnamefont {Kajinami}},
  \bibinfo {author} {\bibfnamefont {Y.}~\bibnamefont {Tabata}}, \bibinfo
  {author} {\bibfnamefont {H.}~\bibnamefont {Nakamura}}, \bibinfo {author}
  {\bibfnamefont {M.}~\bibnamefont {Yoshida}}, \bibinfo {author} {\bibfnamefont
  {M.}~\bibnamefont {Takigawa}},\ and\ \bibinfo {author} {\bibfnamefont
  {I.}~\bibnamefont {Watanabe}},\ }\href@noop {} {\bibfield  {journal}
  {\bibinfo  {journal} {Phys. Rev. B}\ }\textbf {\bibinfo {volume} {81}},\
  \bibinfo {pages} {020401(R)} (\bibinfo {year} {2010})}\BibitemShut {NoStop}%
\bibitem [{\citenamefont {Widmann}\ \emph
  {et~al.}(2016{\natexlab{a}})\citenamefont {Widmann}, \citenamefont {Ruff},
  \citenamefont {G{\"u}nther}, \citenamefont {{Krug von Nidda}}, \citenamefont
  {Lunkenheimer}, \citenamefont {Tsurkan}, \citenamefont {Bord\'{a}cs},
  \citenamefont {K{\'{e}}zsm{\'{a}}rki},\ and\ \citenamefont
  {Loidl}}]{Widmann:2016a}%
  \BibitemOpen
  \bibfield  {author} {\bibinfo {author} {\bibfnamefont {S.}~\bibnamefont
  {Widmann}}, \bibinfo {author} {\bibfnamefont {E.}~\bibnamefont {Ruff}},
  \bibinfo {author} {\bibfnamefont {A.}~\bibnamefont {G{\"u}nther}}, \bibinfo
  {author} {\bibfnamefont {H.-A.}\ \bibnamefont {{Krug von Nidda}}}, \bibinfo
  {author} {\bibfnamefont {P.}~\bibnamefont {Lunkenheimer}}, \bibinfo {author}
  {\bibfnamefont {V.}~\bibnamefont {Tsurkan}}, \bibinfo {author} {\bibfnamefont
  {S.}~\bibnamefont {Bord\'{a}cs}}, \bibinfo {author} {\bibfnamefont
  {I.}~\bibnamefont {K{\'{e}}zsm{\'{a}}rki}},\ and\ \bibinfo {author}
  {\bibfnamefont {A.}~\bibnamefont {Loidl}},\ }\href@noop {} {\bibfield
  {journal} {\bibinfo  {journal} {Philos. Mag.}\ ,\ \bibinfo {pages} {doi:
  10.1080/14786435.2016.1253885}} (\bibinfo {year}
  {2016}{\natexlab{a}})}\BibitemShut {NoStop}%
\bibitem [{\citenamefont {Widmann}\ \emph
  {et~al.}(2016{\natexlab{b}})\citenamefont {Widmann}, \citenamefont
  {G{\"u}nther}, \citenamefont {Ruff}, \citenamefont {Tsurkan}, \citenamefont
  {{Krug von Nidda}}, \citenamefont {Lunkenheimer},\ and\ \citenamefont
  {Loidl}}]{Widmann:2016}%
  \BibitemOpen
  \bibfield  {author} {\bibinfo {author} {\bibfnamefont {S.}~\bibnamefont
  {Widmann}}, \bibinfo {author} {\bibfnamefont {A.}~\bibnamefont
  {G{\"u}nther}}, \bibinfo {author} {\bibfnamefont {E.}~\bibnamefont {Ruff}},
  \bibinfo {author} {\bibfnamefont {V.}~\bibnamefont {Tsurkan}}, \bibinfo
  {author} {\bibfnamefont {H.-A.}\ \bibnamefont {{Krug von Nidda}}}, \bibinfo
  {author} {\bibfnamefont {P.}~\bibnamefont {Lunkenheimer}},\ and\ \bibinfo
  {author} {\bibfnamefont {A.}~\bibnamefont {Loidl}},\ }\href@noop {}
  {\bibfield  {journal} {\bibinfo  {journal} {Phys. Rev. B}\ }\textbf {\bibinfo
  {volume} {94}},\ \bibinfo {pages} {214421} (\bibinfo {year}
  {2016}{\natexlab{b}})}\BibitemShut {NoStop}%
\bibitem [{\citenamefont {Rastogi}\ \emph {et~al.}(1983)\citenamefont
  {Rastogi}, \citenamefont {Berton}, \citenamefont {Chaussy}, \citenamefont
  {Tournier}, \citenamefont {Chevrel},\ and\ \citenamefont
  {Sergent}}]{Rastogi:1983}%
  \BibitemOpen
  \bibfield  {author} {\bibinfo {author} {\bibfnamefont {A.~K.}\ \bibnamefont
  {Rastogi}}, \bibinfo {author} {\bibfnamefont {A.}~\bibnamefont {Berton}},
  \bibinfo {author} {\bibfnamefont {J.}~\bibnamefont {Chaussy}}, \bibinfo
  {author} {\bibfnamefont {R.}~\bibnamefont {Tournier}}, \bibinfo {author}
  {\bibfnamefont {M.~P.~R.}\ \bibnamefont {Chevrel}},\ and\ \bibinfo {author}
  {\bibfnamefont {M.}~\bibnamefont {Sergent}},\ }\href@noop {} {\bibfield
  {journal} {\bibinfo  {journal} {J. Low Temp. Phys.}\ }\textbf {\bibinfo
  {volume} {52}},\ \bibinfo {pages} {539} (\bibinfo {year} {1983})}\BibitemShut
  {NoStop}%
\bibitem [{\citenamefont {Tabata}\ \emph {et~al.}(2010)\citenamefont {Tabata},
  \citenamefont {Kajinami}, \citenamefont {Waki}, \citenamefont {Watanabe},\
  and\ \citenamefont {Nakamura}}]{Tabata:2010}%
  \BibitemOpen
  \bibfield  {author} {\bibinfo {author} {\bibfnamefont {Y.}~\bibnamefont
  {Tabata}}, \bibinfo {author} {\bibfnamefont {Y.}~\bibnamefont {Kajinami}},
  \bibinfo {author} {\bibfnamefont {T.}~\bibnamefont {Waki}}, \bibinfo {author}
  {\bibfnamefont {I.}~\bibnamefont {Watanabe}},\ and\ \bibinfo {author}
  {\bibfnamefont {H.}~\bibnamefont {Nakamura}},\ }\href@noop {} {\bibfield
  {journal} {\bibinfo  {journal} {J. Phys.: Conf. Ser.}\ }\textbf {\bibinfo
  {volume} {225}},\ \bibinfo {pages} {012055} (\bibinfo {year}
  {2010})}\BibitemShut {NoStop}%
\bibitem [{\citenamefont {Bichler}(2010)}]{Bichler:2010}%
  \BibitemOpen
  \bibfield  {author} {\bibinfo {author} {\bibfnamefont {D.}~\bibnamefont
  {Bichler}},\ }\href@noop {} {\bibinfo {type} {{Ph.D.} thesis}},\ \bibinfo
  {school} {Ludwig-Maximilians-Universit{\"a}t M{\"u}nchen} (\bibinfo {year}
  {2010})\BibitemShut {NoStop}%
\bibitem [{\citenamefont {Ikeno}\ \emph {et~al.}(2007)\citenamefont {Ikeno},
  \citenamefont {Nakamura},\ and\ \citenamefont {Kohara}}]{Ikeno:2007}%
  \BibitemOpen
  \bibfield  {author} {\bibinfo {author} {\bibfnamefont {R.}~\bibnamefont
  {Ikeno}}, \bibinfo {author} {\bibfnamefont {H.}~\bibnamefont {Nakamura}},\
  and\ \bibinfo {author} {\bibfnamefont {T.}~\bibnamefont {Kohara}},\
  }\href@noop {} {\bibfield  {journal} {\bibinfo  {journal} {J. Phys.: Condens.
  Matter}\ }\textbf {\bibinfo {volume} {19}},\ \bibinfo {pages} {046206}
  (\bibinfo {year} {2007})}\BibitemShut {NoStop}%
\bibitem [{\citenamefont {Rastogi}\ and\ \citenamefont
  {Wohlfarth}(1987)}]{Rastogi:1987}%
  \BibitemOpen
  \bibfield  {author} {\bibinfo {author} {\bibfnamefont {A.~K.}\ \bibnamefont
  {Rastogi}}\ and\ \bibinfo {author} {\bibfnamefont {E.~P.}\ \bibnamefont
  {Wohlfarth}},\ }\href@noop {} {\bibfield  {journal} {\bibinfo  {journal}
  {Phys. stat. sol. (b)}\ }\textbf {\bibinfo {volume} {142}},\ \bibinfo {pages}
  {569} (\bibinfo {year} {1987})}\BibitemShut {NoStop}%
\bibitem [{\citenamefont {Kawamoto}\ \emph {et~al.}(2016)\citenamefont
  {Kawamoto}, \citenamefont {Higo}, \citenamefont {Tomita}, \citenamefont
  {Suzuki}, \citenamefont {Tian}, \citenamefont {Mochitzuki}, \citenamefont
  {Matsuo}, \citenamefont {Kindo},\ and\ \citenamefont
  {Nakatsuji}}]{Kawamoto:2016}%
  \BibitemOpen
  \bibfield  {author} {\bibinfo {author} {\bibfnamefont {S.}~\bibnamefont
  {Kawamoto}}, \bibinfo {author} {\bibfnamefont {T.}~\bibnamefont {Higo}},
  \bibinfo {author} {\bibfnamefont {T.}~\bibnamefont {Tomita}}, \bibinfo
  {author} {\bibfnamefont {S.}~\bibnamefont {Suzuki}}, \bibinfo {author}
  {\bibfnamefont {Z.~M.}\ \bibnamefont {Tian}}, \bibinfo {author}
  {\bibfnamefont {K.}~\bibnamefont {Mochitzuki}}, \bibinfo {author}
  {\bibfnamefont {A.}~\bibnamefont {Matsuo}}, \bibinfo {author} {\bibfnamefont
  {K.}~\bibnamefont {Kindo}},\ and\ \bibinfo {author} {\bibfnamefont
  {S.}~\bibnamefont {Nakatsuji}},\ }\href@noop {} {\bibfield  {journal}
  {\bibinfo  {journal} {J. Phys.: Conf. Ser.}\ }\textbf {\bibinfo {volume}
  {683}},\ \bibinfo {pages} {012025} (\bibinfo {year} {2016})}\BibitemShut
  {NoStop}%
\bibitem [{\citenamefont {Nakamura}\ \emph {et~al.}(2005)\citenamefont
  {Nakamura}, \citenamefont {Chudo},\ and\ \citenamefont
  {Shiga}}]{Nakamura:2005}%
  \BibitemOpen
  \bibfield  {author} {\bibinfo {author} {\bibfnamefont {H.}~\bibnamefont
  {Nakamura}}, \bibinfo {author} {\bibfnamefont {H.}~\bibnamefont {Chudo}},\
  and\ \bibinfo {author} {\bibfnamefont {M.}~\bibnamefont {Shiga}},\
  }\href@noop {} {\bibfield  {journal} {\bibinfo  {journal} {J. Phys.: Condens.
  Matter}\ }\textbf {\bibinfo {volume} {17}},\ \bibinfo {pages} {6015}
  (\bibinfo {year} {2005})}\BibitemShut {NoStop}%
\bibitem [{\citenamefont {Leonov}\ and\ \citenamefont
  {K{\'{e}}zsm{\'{a}}rki}(2017{\natexlab{a}})}]{Leonov:2017}%
  \BibitemOpen
  \bibfield  {author} {\bibinfo {author} {\bibfnamefont {A.~O.}\ \bibnamefont
  {Leonov}}\ and\ \bibinfo {author} {\bibfnamefont {I.}~\bibnamefont
  {K{\'{e}}zsm{\'{a}}rki}},\ }\href@noop {} {\bibfield  {journal} {\bibinfo
  {journal} {Phys. Rev. B}\ }\textbf {\bibinfo {volume} {96}},\ \bibinfo
  {pages} {014423} (\bibinfo {year} {2017}{\natexlab{a}})}\BibitemShut
  {NoStop}%
\bibitem [{\citenamefont {Leonov}\ and\ \citenamefont
  {K{\'{e}}zsm{\'{a}}rki}(2017{\natexlab{b}})}]{Leonov:2017a}%
  \BibitemOpen
  \bibfield  {author} {\bibinfo {author} {\bibfnamefont {A.~O.}\ \bibnamefont
  {Leonov}}\ and\ \bibinfo {author} {\bibfnamefont {I.}~\bibnamefont
  {K{\'{e}}zsm{\'{a}}rki}},\ }\href@noop {} {\bibfield  {journal} {\bibinfo
  {journal} {Phys. Rev. B}\ }\textbf {\bibinfo {volume} {96}},\ \bibinfo
  {pages} {214413} (\bibinfo {year} {2017}{\natexlab{b}})}\BibitemShut
  {NoStop}%
\bibitem [{\citenamefont {Ruff}\ \emph {et~al.}(2015)\citenamefont {Ruff},
  \citenamefont {Widmann}, \citenamefont {Lunkenheimer}, \citenamefont
  {Tsurkan}, \citenamefont {Bord\'{a}cs}, \citenamefont
  {K{\'{e}}zsm{\'{a}}rki},\ and\ \citenamefont {Loidl}}]{Ruff:2015}%
  \BibitemOpen
  \bibfield  {author} {\bibinfo {author} {\bibfnamefont {E.}~\bibnamefont
  {Ruff}}, \bibinfo {author} {\bibfnamefont {S.}~\bibnamefont {Widmann}},
  \bibinfo {author} {\bibfnamefont {P.}~\bibnamefont {Lunkenheimer}}, \bibinfo
  {author} {\bibfnamefont {V.}~\bibnamefont {Tsurkan}}, \bibinfo {author}
  {\bibfnamefont {S.}~\bibnamefont {Bord\'{a}cs}}, \bibinfo {author}
  {\bibfnamefont {I.}~\bibnamefont {K{\'{e}}zsm{\'{a}}rki}},\ and\ \bibinfo
  {author} {\bibfnamefont {A.}~\bibnamefont {Loidl}},\ }\href@noop {}
  {\bibfield  {journal} {\bibinfo  {journal} {Sci. Adv.}\ ,\ \bibinfo {pages}
  {1:e1500916}} (\bibinfo {year} {2015})}\BibitemShut {NoStop}%
\bibitem [{\citenamefont {Xu}\ and\ \citenamefont {Xiang}(2015)}]{Xu:2015}%
  \BibitemOpen
  \bibfield  {author} {\bibinfo {author} {\bibfnamefont {K.}~\bibnamefont
  {Xu}}\ and\ \bibinfo {author} {\bibfnamefont {H.~J.}\ \bibnamefont {Xiang}},\
  }\href@noop {} {\bibfield  {journal} {\bibinfo  {journal} {Phys. Rev. B}\
  }\textbf {\bibinfo {volume} {92}},\ \bibinfo {pages} {121112(R)} (\bibinfo
  {year} {2015})}\BibitemShut {NoStop}%
\bibitem [{\citenamefont {Zhang}\ \emph {et~al.}(2017)\citenamefont {Zhang},
  \citenamefont {Wang}, \citenamefont {Yang}, \citenamefont {Xia},
  \citenamefont {Lu},\ and\ \citenamefont {Zhu}}]{Zhang:2017}%
  \BibitemOpen
  \bibfield  {author} {\bibinfo {author} {\bibfnamefont {J.~T.}\ \bibnamefont
  {Zhang}}, \bibinfo {author} {\bibfnamefont {J.~L.}\ \bibnamefont {Wang}},
  \bibinfo {author} {\bibfnamefont {X.~Q.}\ \bibnamefont {Yang}}, \bibinfo
  {author} {\bibfnamefont {W.~S.}\ \bibnamefont {Xia}}, \bibinfo {author}
  {\bibfnamefont {X.~M.}\ \bibnamefont {Lu}},\ and\ \bibinfo {author}
  {\bibfnamefont {J.~S.}\ \bibnamefont {Zhu}},\ }\href@noop {} {\bibfield
  {journal} {\bibinfo  {journal} {Phys. Rev. B}\ }\textbf {\bibinfo {volume}
  {95}},\ \bibinfo {pages} {085136} (\bibinfo {year} {2017})}\BibitemShut
  {NoStop}%
\bibitem [{\citenamefont {Nikolaev}\ and\ \citenamefont
  {Solovyev}(2019)}]{Nikolaev:2019}%
  \BibitemOpen
  \bibfield  {author} {\bibinfo {author} {\bibfnamefont {S.~A.}\ \bibnamefont
  {Nikolaev}}\ and\ \bibinfo {author} {\bibfnamefont {I.~V.}\ \bibnamefont
  {Solovyev}},\ }\href@noop {} {\bibfield  {journal} {\bibinfo  {journal}
  {Phys. Rev. B}\ }\textbf {\bibinfo {volume} {99}},\ \bibinfo {pages}
  {100401(R)} (\bibinfo {year} {2019})}\BibitemShut {NoStop}%
\bibitem [{\citenamefont {Zhang}\ \emph {et~al.}(2019)\citenamefont {Zhang},
  \citenamefont {Chen}, \citenamefont {Barone}, \citenamefont {Yamauchi},
  \citenamefont {Dong},\ and\ \citenamefont {Picozzi}}]{Zhang:2019}%
  \BibitemOpen
  \bibfield  {author} {\bibinfo {author} {\bibfnamefont {H.-M.}\ \bibnamefont
  {Zhang}}, \bibinfo {author} {\bibfnamefont {J.}~\bibnamefont {Chen}},
  \bibinfo {author} {\bibfnamefont {P.}~\bibnamefont {Barone}}, \bibinfo
  {author} {\bibfnamefont {K.}~\bibnamefont {Yamauchi}}, \bibinfo {author}
  {\bibfnamefont {S.}~\bibnamefont {Dong}},\ and\ \bibinfo {author}
  {\bibfnamefont {S.}~\bibnamefont {Picozzi}},\ }\href@noop {} {\bibfield
  {journal} {\bibinfo  {journal} {Phys. Rev. B}\ }\textbf {\bibinfo {volume}
  {99}},\ \bibinfo {pages} {214427} (\bibinfo {year} {2019})}\BibitemShut
  {NoStop}%
\bibitem [{\citenamefont {{D. A. Kitchaev and E. C. Schueller and A. Van der
  Ven}}(2019)}]{Kitchaev:2019}%
  \BibitemOpen
  \bibfield  {author} {\bibinfo {author} {\bibnamefont {{D. A. Kitchaev and E.
  C. Schueller and A. Van der Ven}}},\ }\href@noop {} {\bibfield  {journal}
  {\bibinfo  {journal} {Phys. Rev. B}\ } \textbf {\bibinfo {volume}
  {101}},\ \bibinfo {pages} {054409} (\bibinfo {year}
  {2020})}\BibitemShut {NoStop}%
\bibitem [{\citenamefont {Katukuri}\ \emph {et~al.}(2014)\citenamefont
  {Katukuri}, \citenamefont {Nishimoto}, \citenamefont {Yushankhai},
  \citenamefont {Stoyanova}, \citenamefont {Kandpal}, \citenamefont {Choi},
  \citenamefont {Coldea}, \citenamefont {Rousochatzakis}, \citenamefont
  {Hozoi},\ and\ \citenamefont {van~den Brink}}]{Katukuri:2014}%
  \BibitemOpen
  \bibfield  {author} {\bibinfo {author} {\bibfnamefont {V.~M.}\ \bibnamefont
  {Katukuri}}, \bibinfo {author} {\bibfnamefont {S.}~\bibnamefont {Nishimoto}},
  \bibinfo {author} {\bibfnamefont {V.}~\bibnamefont {Yushankhai}}, \bibinfo
  {author} {\bibfnamefont {A.}~\bibnamefont {Stoyanova}}, \bibinfo {author}
  {\bibfnamefont {H.}~\bibnamefont {Kandpal}}, \bibinfo {author} {\bibfnamefont
  {S.}~\bibnamefont {Choi}}, \bibinfo {author} {\bibfnamefont {R.}~\bibnamefont
  {Coldea}}, \bibinfo {author} {\bibfnamefont {I.}~\bibnamefont
  {Rousochatzakis}}, \bibinfo {author} {\bibfnamefont {L.}~\bibnamefont
  {Hozoi}},\ and\ \bibinfo {author} {\bibfnamefont {J.}~\bibnamefont {van~den
  Brink}},\ }\href@noop {} {\bibfield  {journal} {\bibinfo  {journal} {New J.
  Phys.}\ }\textbf {\bibinfo {volume} {16}},\ \bibinfo {pages} {013056}
  (\bibinfo {year} {2014})}\BibitemShut {NoStop}%
\bibitem [{\citenamefont {Yadav}\ \emph {et~al.}(2016)\citenamefont {Yadav},
  \citenamefont {Bogdanov}, \citenamefont {Katukuri}, \citenamefont
  {Nishimoto}, \citenamefont {van~den Brink},\ and\ \citenamefont
  {Hozoi}}]{Yadav:2016}%
  \BibitemOpen
  \bibfield  {author} {\bibinfo {author} {\bibfnamefont {R.}~\bibnamefont
  {Yadav}}, \bibinfo {author} {\bibfnamefont {N.~A.}\ \bibnamefont {Bogdanov}},
  \bibinfo {author} {\bibfnamefont {V.~M.}\ \bibnamefont {Katukuri}}, \bibinfo
  {author} {\bibfnamefont {S.}~\bibnamefont {Nishimoto}}, \bibinfo {author}
  {\bibfnamefont {J.}~\bibnamefont {van~den Brink}},\ and\ \bibinfo {author}
  {\bibfnamefont {L.}~\bibnamefont {Hozoi}},\ }\href@noop {} {\bibfield
  {journal} {\bibinfo  {journal} {Sci. Rep.}\ }\textbf {\bibinfo {volume}
  {6}},\ \bibinfo {pages} {37925} (\bibinfo {year} {2016})}\BibitemShut
  {NoStop}%
\bibitem [{\citenamefont {Hozoi}\ \emph {et~al.}(2011)\citenamefont {Hozoi},
  \citenamefont {Siurakshina}, \citenamefont {Fulde},\ and\ \citenamefont
  {van~den Brink}}]{Hozoi:2011}%
  \BibitemOpen
  \bibfield  {author} {\bibinfo {author} {\bibfnamefont {L.}~\bibnamefont
  {Hozoi}}, \bibinfo {author} {\bibfnamefont {L.}~\bibnamefont {Siurakshina}},
  \bibinfo {author} {\bibfnamefont {P.}~\bibnamefont {Fulde}},\ and\ \bibinfo
  {author} {\bibfnamefont {J.}~\bibnamefont {van~den Brink}},\ }\href@noop {}
  {\bibfield  {journal} {\bibinfo  {journal} {Sci. Rep.}\ }\textbf {\bibinfo
  {volume} {1}},\ \bibinfo {pages} {65} (\bibinfo {year} {2011})}\BibitemShut
  {NoStop}%
\bibitem [{\citenamefont {Kuzmenko}(2005)}]{Kuzmenko:2005}%
  \BibitemOpen
  \bibfield  {author} {\bibinfo {author} {\bibfnamefont {A.~B.}\ \bibnamefont
  {Kuzmenko}},\ }\href@noop {} {\bibfield  {journal} {\bibinfo  {journal} {Rev.
  Sci. Instr.}\ }\textbf {\bibinfo {volume} {76}},\ \bibinfo {pages} {083108}
  (\bibinfo {year} {2005})}\BibitemShut {NoStop}%
\bibitem [{\citenamefont {Kuzmenko}(2018)}]{Kuzmenko:2018}%
  \BibitemOpen
  \bibfield  {author} {\bibinfo {author} {\bibfnamefont {A.~B.}\ \bibnamefont
  {Kuzmenko}},\ }\href@noop {} {\bibinfo {title} {{RefFIT v. 1.3.05}}}
  (\bibinfo {year} {2018}),\ \bibinfo {note} {https://reffit.ch/}\BibitemShut
  {NoStop}%
\bibitem [{\citenamefont {Reschke}\ \emph {et~al.}(2017)\citenamefont
  {Reschke}, \citenamefont {Mayr}, \citenamefont {Wang}, \citenamefont
  {Lunkenheimer}, \citenamefont {Li}, \citenamefont {Szaller}, \citenamefont
  {Bord\'{a}cs}, \citenamefont {K{\'{e}}zsm{\'{a}}rki}, \citenamefont
  {Tsurkan},\ and\ \citenamefont {Loidl}}]{Reschke:2017}%
  \BibitemOpen
  \bibfield  {author} {\bibinfo {author} {\bibfnamefont {S.}~\bibnamefont
  {Reschke}}, \bibinfo {author} {\bibfnamefont {F.}~\bibnamefont {Mayr}},
  \bibinfo {author} {\bibfnamefont {Z.}~\bibnamefont {Wang}}, \bibinfo {author}
  {\bibfnamefont {P.}~\bibnamefont {Lunkenheimer}}, \bibinfo {author}
  {\bibfnamefont {W.}~\bibnamefont {Li}}, \bibinfo {author} {\bibfnamefont
  {D.}~\bibnamefont {Szaller}}, \bibinfo {author} {\bibfnamefont
  {S.}~\bibnamefont {Bord\'{a}cs}}, \bibinfo {author} {\bibfnamefont
  {I.}~\bibnamefont {K{\'{e}}zsm{\'{a}}rki}}, \bibinfo {author} {\bibfnamefont
  {V.}~\bibnamefont {Tsurkan}},\ and\ \bibinfo {author} {\bibfnamefont
  {A.}~\bibnamefont {Loidl}},\ }\href@noop {} {\bibfield  {journal} {\bibinfo
  {journal} {Phys. Rev. B}\ }\textbf {\bibinfo {volume} {96}},\ \bibinfo
  {pages} {144302} (\bibinfo {year} {2017})}\BibitemShut {NoStop}%
\bibitem [{\citenamefont {Hlinka}\ \emph {et~al.}(2016)\citenamefont {Hlinka},
  \citenamefont {Borodavka}, \citenamefont {Rafalovskyi}, \citenamefont
  {Docekalova}, \citenamefont {Pokorny}, \citenamefont {Gregora}, \citenamefont
  {Tsurkan}, \citenamefont {Nakamura}, \citenamefont {Mayr}, \citenamefont
  {Kuntscher}, \citenamefont {Loidl}, \citenamefont {Bord\'{a}cs},
  \citenamefont {Szaller}, \citenamefont {Lee}, \citenamefont {Lee},\ and\
  \citenamefont {K{\'{e}}zsm{\'{a}}rki}}]{Hlinka:2016}%
  \BibitemOpen
  \bibfield  {author} {\bibinfo {author} {\bibfnamefont {J.}~\bibnamefont
  {Hlinka}}, \bibinfo {author} {\bibfnamefont {F.}~\bibnamefont {Borodavka}},
  \bibinfo {author} {\bibfnamefont {I.}~\bibnamefont {Rafalovskyi}}, \bibinfo
  {author} {\bibfnamefont {Z.}~\bibnamefont {Docekalova}}, \bibinfo {author}
  {\bibfnamefont {J.}~\bibnamefont {Pokorny}}, \bibinfo {author} {\bibfnamefont
  {I.}~\bibnamefont {Gregora}}, \bibinfo {author} {\bibfnamefont
  {V.}~\bibnamefont {Tsurkan}}, \bibinfo {author} {\bibfnamefont
  {H.}~\bibnamefont {Nakamura}}, \bibinfo {author} {\bibfnamefont
  {F.}~\bibnamefont {Mayr}}, \bibinfo {author} {\bibfnamefont {C.~A.}\
  \bibnamefont {Kuntscher}}, \bibinfo {author} {\bibfnamefont {A.}~\bibnamefont
  {Loidl}}, \bibinfo {author} {\bibfnamefont {S.}~\bibnamefont {Bord\'{a}cs}},
  \bibinfo {author} {\bibfnamefont {D.}~\bibnamefont {Szaller}}, \bibinfo
  {author} {\bibfnamefont {H.-J.}\ \bibnamefont {Lee}}, \bibinfo {author}
  {\bibfnamefont {J.~H.}\ \bibnamefont {Lee}},\ and\ \bibinfo {author}
  {\bibfnamefont {I.}~\bibnamefont {K{\'{e}}zsm{\'{a}}rki}},\ }\href@noop {}
  {\bibfield  {journal} {\bibinfo  {journal} {Phys. Rev. B}\ }\textbf {\bibinfo
  {volume} {94}},\ \bibinfo {pages} {060104(R)} (\bibinfo {year}
  {2016})}\BibitemShut {NoStop}%
\bibitem [{\citenamefont {Guiot}\ \emph {et~al.}(2013)\citenamefont {Guiot},
  \citenamefont {Cario}, \citenamefont {Janod}, \citenamefont {Corraze},
  \citenamefont {{Ta Phuoc}}, \citenamefont {Rozenberg}, \citenamefont
  {Stoliar}, \citenamefont {Cren},\ and\ \citenamefont
  {Roditchev}}]{Guiot:2013}%
  \BibitemOpen
  \bibfield  {author} {\bibinfo {author} {\bibfnamefont {V.}~\bibnamefont
  {Guiot}}, \bibinfo {author} {\bibfnamefont {L.}~\bibnamefont {Cario}},
  \bibinfo {author} {\bibfnamefont {E.}~\bibnamefont {Janod}}, \bibinfo
  {author} {\bibfnamefont {B.}~\bibnamefont {Corraze}}, \bibinfo {author}
  {\bibfnamefont {V.}~\bibnamefont {{Ta Phuoc}}}, \bibinfo {author}
  {\bibfnamefont {M.}~\bibnamefont {Rozenberg}}, \bibinfo {author}
  {\bibfnamefont {P.}~\bibnamefont {Stoliar}}, \bibinfo {author} {\bibfnamefont
  {T.}~\bibnamefont {Cren}},\ and\ \bibinfo {author} {\bibfnamefont
  {D.}~\bibnamefont {Roditchev}},\ }\href@noop {} {\bibfield  {journal}
  {\bibinfo  {journal} {Nat. Commun.}\ }\textbf {\bibinfo {volume} {4}},\
  \bibinfo {pages} {1722} (\bibinfo {year} {2013})}\BibitemShut {NoStop}%
\bibitem [{\citenamefont {Bichler}\ and\ \citenamefont
  {Johrendt}(2011)}]{Bichler:2011}%
  \BibitemOpen
  \bibfield  {author} {\bibinfo {author} {\bibfnamefont {D.}~\bibnamefont
  {Bichler}}\ and\ \bibinfo {author} {\bibfnamefont {D.}~\bibnamefont
  {Johrendt}},\ }\href@noop {} {\bibfield  {journal} {\bibinfo  {journal}
  {Chem. Mater.}\ }\textbf {\bibinfo {volume} {23}},\ \bibinfo {pages} {3014}
  (\bibinfo {year} {2011})}\BibitemShut {NoStop}%
\bibitem [{\citenamefont {Guiot}\ \emph {et~al.}(2011)\citenamefont {Guiot},
  \citenamefont {Janod}, \citenamefont {Corraze},\ and\ \citenamefont
  {Cario}}]{Guiot:2011}%
  \BibitemOpen
  \bibfield  {author} {\bibinfo {author} {\bibfnamefont {V.}~\bibnamefont
  {Guiot}}, \bibinfo {author} {\bibfnamefont {E.}~\bibnamefont {Janod}},
  \bibinfo {author} {\bibfnamefont {B.}~\bibnamefont {Corraze}},\ and\ \bibinfo
  {author} {\bibfnamefont {L.}~\bibnamefont {Cario}},\ }\href@noop {}
  {\bibfield  {journal} {\bibinfo  {journal} {Chem. Mater.}\ }\textbf {\bibinfo
  {volume} {23}},\ \bibinfo {pages} {2611} (\bibinfo {year}
  {2011})}\BibitemShut {NoStop}%
\bibitem [{\citenamefont {Lee}\ \emph {et~al.}(2019)\citenamefont {Lee},
  \citenamefont {Jeong}, \citenamefont {Sim}, \citenamefont {Yoon},
  \citenamefont {Ryee},\ and\ \citenamefont {Han}}]{Lee:2019}%
  \BibitemOpen
  \bibfield  {author} {\bibinfo {author} {\bibfnamefont {H.}~\bibnamefont
  {Lee}}, \bibinfo {author} {\bibfnamefont {M.~Y.}\ \bibnamefont {Jeong}},
  \bibinfo {author} {\bibfnamefont {J.-H.}\ \bibnamefont {Sim}}, \bibinfo
  {author} {\bibfnamefont {H.}~\bibnamefont {Yoon}}, \bibinfo {author}
  {\bibfnamefont {S.}~\bibnamefont {Ryee}},\ and\ \bibinfo {author}
  {\bibfnamefont {M.~J.}\ \bibnamefont {Han}},\ }\href@noop {} {\bibfield
  {journal} {\bibinfo  {journal} {EPL}\ }\textbf {\bibinfo {volume} {125}},\
  \bibinfo {pages} {47005} (\bibinfo {year} {2019})}\BibitemShut {NoStop}%
\end{thebibliography}
\end{document}